\documentclass[twoside]{article}
\usepackage{a4wide}
\usepackage{amsmath}
\usepackage{amsfonts}
\usepackage{amssymb}
\usepackage{pstricks,pst-node,pst-text,pst-3d}
\usepackage{times}
\usepackage{graphics}
\def\date{Jan 1, 1905}
\newcommand{\ed}{\end{document}}
\renewcommand{\theequation}{\arabic{section}-\arabic{equation}}
\newcounter{mycnt}[section]

\def\one{\hbox{{1}\kern-.25em\hbox{l}}}

\def\!{\kern -0.15ex}

\begin{document}
\title{Path integral formulation and Feynman rules for phylogenetic branching 
models.}
\author{P D Jarvis$^{\dagger}$, 
J D Bashford$^{*}$
and J G Sumner}
{ {\renewcommand{\thefootnote}{\fnsymbol{footnote}}
\footnotetext{\kern-15.3pt AMS Subject Classification:
05C05; 
60J80; 
92D20; 
81S40  
. \\ $^{*}$Australian Postdoctoral Fellow \\
$^{\dagger}$Alexander von Humboldt Fellow

}
}}

\maketitle

\begin{abstract}
A dynamical picture of phylogenetic evolution is given in terms of
Markov models on a state space, comprising joint probability
distributions for character types of taxonomic classes.  Phylogenetic
branching is a process which augments the number of taxa under
consideration, and hence the rank of the underlying joint probability
state tensor.  We point out the combinatorial necessity for a
second-quantised, or Fock space setting, incorporating discrete
counting labels for taxa and character types, to allow for a description in the number
basis.  Rate operators describing both time evolution
without branching, and also phylogenetic branching events, are
identified. A detailed development of these ideas is given, using standard
transcriptions from the microscopic formulation of nonequilibrium
reaction-diffusion or birth-death processes.  These give the relations
between stochastic rate matrices, the matrix elements of the
corresponding evolution operators representing them, and the integral
kernels needed to implement these as path integrals.  
The `free'  theory (without branching) is solved, and the correct trilinear
`interaction'  terms (representing branching events) are presented. 
The full model is developed in perturbation theory via the derivation of 
explicit Feynman rules which establish that
the probabilities (pattern frequencies of leaf colourations)
arising as matrix elements of the time evolution operator
are identical with those computed via the standard analysis. 
Simple examples (phylogenetic trees with 2 or 3 leaves), are
discussed in detail. Further implications for the work are briefly considered
including the role of time reparametrisation covariance.

\end{abstract}
\pagebreak

\section{Introduction and background}
\label{sec:Intro}
The use of Markov models of change to taxonomic character probability
distributions is a standard technique for describing mutations, and
for inferring ancestral relationships between taxa.  A general
stochastic framework for phylogenetic branching models is as follows. 
By assumption, different `taxonomic units' are identified, and
classified by a set of defining characteristics: based on
morphological features for example, or on sequence data for a
particular gene or protein say.  To each taxon is ascribed a
probability density on the set of characters, and it is the task of
phylogenetic reconstruction to infer ancestral relationships within a
group of taxa, given observed pattern frequencies for characters
amongst the taxa (definitions are given later in the text) .  In such 
phylogenetic reconstruction, the Markov chain model describing the 
stochastic evolution of characters is
extended appropriately to encompass `branching' where the number of
taxa is augmented as new taxonomic types evolve (for example by
speciation or gene duplication), from an initial single
progenitor, through to the final number of types under study.
For details of the subject, incuding overviews 
of applications, current problems and new directions, we refer to 
recent textbooks, for example \cite{SempleSteel2003,Felsenstein2004}.

In recent work
\cite{BashfordJarvis2001,BashfordJarvisSumnerSteel2004,SumnerJarvis2004,SumnerBashfordJarvis2005}
it has been pointed out that a fruitful approach to phylogenetic
analysis is afforded by taking the formal perspective of multilinear
tensor algebra familiar from physical systems.  For example, in the
analysis of symmetry properties (of the Markov rate matrix, and of the
branching process) it is natural to consider \textit{continuous} Lie
transformation groups acting on the tensor spaces, and the associated
representation theory \cite{BashfordJarvisSumnerSteel2004}. 
Furthermore, a remarkable analogy between branching processes (where
the technical constraint of local conditional independence
\cite{SempleSteel2003} is imposed) and state entanglement in quantum
physics has been noted \cite{SumnerJarvis2004}.  In particular, for 2
characters (equivalent to single qubit (2 state) systems in quantum
physics) the well-known $\log \det$ distance measure for 2 taxa is
essentially the \textit{concurrence} (for 2 qubits, related to the 
von Neumann
entropy of a partial density operator); equally the \textit{tangle}
(an entanglement measure for 3 qubits) has been proposed as a useful
measure of distance for 3 taxa in the two character case, and the
analysis of its properties in the phylogenetic context has been
initiated \cite{SumnerJarvis2004,SumnerJarvis2005}.  Further applications of classical invariant
theory for phylogenetic analysis are
developed in \cite{SumnerBashfordJarvis2005}.

In the letter \cite{BashfordJarvis2001} it was argued that a
further useful perspective on phylogenetics, again inspired by
physics, can be gained by interpreting `branching' in the model as a
linear operator which \textit{augments} the rank of the tensor
corresponding to the joint probability distribution of character types
(see also \cite{BashfordJarvisSumnerSteel2004,SumnerJarvis2004}).  In
order to regard the entire model, including especially the time 
development represented by the branching dynamics, in
a uniform way, it is natural to seek a setting in multilinear algebra
where the linear space describing state probabilities for taxa can be
lifted to an appropriate free algebra in the sense of tensor products,
or `Fock space' in physical language, so that the linear `branching
operator' has a uniform (extended) domain of definition.  Possible
interaction terms representing this operator, corresponding to
phylogenetic branching events, can then readily be implemented in the
language of second quantization as shown in 
\cite{BashfordJarvis2001}. 
Although formal, the transcription to physical language provided does
indeed establish that the entire Markov branching model can be
regarded as a standard Markov chain, but with dynamics on a suitably
extended state space -- a fact not noted explicitly before.  With
closed form expressions for the probabilities in hand, it may also be
possible to investigate these from various analytical viewpoints not
accessible hitherto.  Moreover, the physical language is quite
flexible, and may suggest useful insights into the models as well as
generalisations.

In the present paper  a
further step towards such analytical investigations and
generalisations is taken, in that the second-quantised framework is
transcribed into the language of path integrals.  The dynamical quantities of 
interest become phylogenetic `path'
variables (or `classical fields'), defined over a discrete spatial
lattice.  Time evolution of the system is developed in
perturbation theory, yielding standard probabilities as convolutions
of the appropriate kernel with the initial probability distribution,
that is, as matrix elements of the evolution operator.  Similar models of 
reaction-diffusion or birth-death processes have
been extensively investigated \cite{Doi1976a,Doi1976b,Peliti1985,
AlcarazDrozHenkelRittenberg1994} so that there is a wealth of
technical experience within this approach, and possibilities for
generalisation.  These introductory comments are supplemented in the
conclusions by further discussion of possible applications (see
summary below).

The outline of the paper is as follows\footnote{For the benefit of 
readers unfamiliar with the subject-matter, technicalities in 
various sections below are treated as fully as possible.}. In 
\S \ref{sec:TensorMethods}
below, we give an analysis of standard accounts of phylogenetic
processes (as used for example in analyses for
inferring ancestral trees) to justify our claim that a multilinear
tensor description is appropriate, and equivalent to the usual
approach.  A standard notation is introduced including the
branching or `splitting' operator whose properties are discussed.  In 
\S
\ref{sec:Fock} the rate operator and the branching operator are
re-formulated as interaction terms in an extended time evolution over
Fock space.  Attention is given to the `copy space' needed to identify
taxa -- both for the observed taxa (leaves) and ancestral stages
(`internal' edges of the phylogenetic tree) -- and it is argued that for
models with $L$ distinguished leaves, a $2^{L}$-dimensional `label'
space is needed.  Label summations suggest a natural
identification with the `momentum' space for
periodic functions over a hypercubic `spatial' lattice in $L$
dimensions (with $2^{L}$ nodes in the unit cell), leading to the
possibility of viewing the system dually in `position' space.  \S 
\ref{sec:PathIntRev} gives a
brief pedagogical review of standard path integral techniques as 
applied for the
analysis of nonequilibrium reaction-diffusion systems in a microscopic
approach.  In \S \ref{sec:FreePathInt} these ingredients are
synthesised in a path integral formulation for a `free' phylogenetic
system, that is a collection of up to $L$ taxa with no phylogenetic
association (not necessarily in a stationary state).  It is shown that
the abstract dynamics, represented by the evolution kernel of the
system in the path integral approach which is formulated and derived
explicitly, does indeed make the system evolve in a standard way
according to a continuous Markov branching process.  In \S
\ref{sec:Interactions} the question of the branching operator is
resumed, and plausible interaction terms (and corresponding normal
kernels) are introduced in the path integral language.  It is shown in
simple examples (trees with 2 or 3 leaves) that, in both the operator 
and path integral language picture, the probabilities arising as matrix 
elements from the dynamics of the model
are identical to those computed in standard likelihood
analyses for inferring phylogenetic trees. This is borne out in the appendix, \S A,
where formal Feynman rules are derived directly from perturbation theory,
and which can immediately be seen to encode the usual sum over 
extended leaf colourations
presentations. The conclusions, \S
\ref{sec:Concl}, reiterate the main points of the paper and further
implications and applications of our work are briefly discussed.
In particular, we comment on the role of the group of 
time reparametrisations (diffeomorphisms),
in the issue of assigning `true' historical time to phylogenetic events.
\section{Tensor methods and stochastic models of phylogenetic 
branching}
\label{sec:TensorMethods}
It is usual to pose the standard stochastic model of phylogenetics by
stating transition probabilities
\cite{Chang1996,SteelHendyPenny1998,SempleSteel2003,Felsenstein2004}.  It is, however,
possible to present the same system in an abstract multilinear tensor
setting.  Our philosophy here is similar to that of
\cite{McCullagh1987} (see also \cite{Diaconis1985}).  In such a
formulation, the evolution of the phylogenetic system is represented 
by a group
action on a tensor product space, with the branching structure
formalized by the introduction of linear `splitting' operators which
increment the rank of the tensor space.  As
pointed out in the introduction, this basis-independent description 
has
many advantages, prompting the investigation of the rich algebraic
structure of the system.  The door is opened to the discussion of 
symmetry
groups and subgroups, representation theory and diagonalization, the
differential structure of the rate operators and orbit classes of
their action, and the ring structure of invariant functions (see
\cite{BashfordJarvisSumnerSteel2004,SumnerJarvis2004,SumnerBashfordJarvis2005}).

Introduce a set, $\mathcal{K}$, which consists of $K$ discrete
elements, conventionally labelled by the integers 
$\{0,1,2,...,K-1\}$. 
Consider a system consisting of $N$ `samples' to each of which can be
attributed one of $K$ distinct characters.  Associated with such a
system we have the set of frequencies
\begin{align}
    \widehat{p}^{\alpha}:&=\frac{\text{total number of occurrences
     of character }\alpha}{N}, \quad \alpha =0,1,...,K-1.\nonumber
\end{align}
In particular we are interested in the character frequencies occurring
in the genome of a given taxon.  The archetypical example is that of
the DNA sequence, where the `samples' are sites, and with four
characters $\{A,G,C,T\}$, but it is, of course, possible to envisage
the use of other character sets derived from the molecular data, so
$K$ is left general in this discussion.  Examples include the amino
acids ($K=20$), codons ($K=64$) or instead of nucleic acid bases 
themselves, a binary pyrimidine/purine $Y/R$ classification of them
($K=2$).  For practical purposes, the usual practice is to take one
particular gene of an organism as being the representative for the
taxon class, although it would be possible to sample a whole genome or
set of genomes and calculate the character frequencies across that
set, and take those frequencies as the representative for that taxon. 
Practical considerations aside, we proceed to model the time evolution
of these frequencies stochastically.

Introduce a random variable $X$ which takes on values in
$\mathcal{K}$.  It is necessary to define a set of time-dependent
probabilities which are the theoretical limit
\begin{align}\label{eq:singleprobs}
    p^{\alpha}(t):=\,&\mathbb{P}(X=\alpha,t)=
    \lim_{N\rightarrow\infty}\widehat{p}^{\alpha}(t).
\end{align}
The stochastic evolution of the probablities is assumed to satisfy
the continuous time Markov property, that is that the state at time 
$t$ depends only on the immediately preceding state at time $t-\delta 
t$ say, and hence   
\begin{align}
p^{\alpha}(t)=&\,\sum_{\beta\in {\mathcal K}} 
\mathbb{P}(X=\alpha,t|X=\beta,t-\delta t) p^{\beta}(t-\delta t),
\end{align}  
which in turn implies, assuming linearity and differentiability,
\begin{align}
    \frac{d}{dt}{p}^{\alpha}(t)=&\,\sum_{\beta\in {\mathcal 
K}}\lim_{\delta t\rightarrow 0}
   \frac{\mathbb{P}(X=\alpha,t +\delta 
t|X=\beta,t)-{\delta^{\alpha}}_ {\beta}}
   {\delta 
   t}p^{\beta}(t).
\end{align}
We define the (time dependent) rate matrix
\begin{align}
{R^{\alpha}}_{\beta}(t)=&\lim_{\delta t\rightarrow 
0}\frac{\mathbb{P}(X=\alpha,t+\delta 
t|X=\beta,t)-{\delta^{\alpha}}_{ \beta}}{\delta t}.
\end{align}
To preserve reality of the probabilities and the property 
$\sum_{\alpha}
p^{\alpha}(t)=1$ for all $t$ it follows that $R$ is a real-valued 
zero column
sum matrix.  In order to preserve positivity of the probabilities it
must also be the case that for all $t$
\begin{align}
{R^{\alpha}}_{\beta}(t) & \geq 0,\quad  \forall \, \alpha \neq 
\beta;\qquad 
{R^{\alpha}}_{\alpha}(t) \leq 0 \quad \mbox{(no sum)}.
\end{align}
For a homogeneous model the rate matrix is assumed to be time
\textit{independent}, with solution
\begin{align}
    \label{eq:singlestoch}
      p^{\alpha}(t) &= 
\sum_{\beta}{M^{\alpha}}_{\beta}(t)p^{\beta}(0), 
     \qquad 
     {M^{\alpha}}_{\beta}(t) ={[e^{Rt}]^{\alpha}}_{\beta},
\end{align}
where $\exp({Rt})$ is calculated using matrix multiplication.

Phylogenetics is concerned with deriving the past evolutionary
relationships of multiple taxa.  As already mentioned, the modern
approach is to compare the genomes of the taxa.  An essential part of
the analysis is the ability to align the genomes of distinct taxa
successfully.  (The possibility or otherwise of such alignment is not
discussed here).  Having aligned the genomes it is possible to
calculate \textit{pattern} frequences.  These patterns are read off
`vertically' across the aligned sequences.  The data is then
\begin{align}
    \widehat{P}^{\alpha_1\alpha_2...\alpha_L}:=\,&\frac{\text{total 
number of occurrences 
    of pattern }\alpha_1\alpha_2...\alpha_L}{N}, \nonumber\\
    \alpha_1,\alpha_2,...,\alpha_L=&0,...,K-1.\nonumber
\end{align}
Introduce random variables $X_1,X_2,...,X_L$ each of which takes on
values in the individual character spaces $\mathcal{K}$, and
$X=(X_1X_2\ldots X_L)$ which takes on values in the $L$-component
character space $\mathcal{K}\times\mathcal{K}\times \cdots 
\times\mathcal{K}$. We model these pattern frequencies by again 
defining a set of time-dependent 
probabilities which are the theoretical limit
\begin{align}
    P^{\alpha_1\alpha_2...\alpha_L}(t):=&\,\mathbb{P}(X=\alpha_1\alpha_2...\alpha_L,t)=
        \lim_{N\rightarrow\infty}\widehat{P}^{\alpha_1\alpha_2...\alpha_L}(t).\nonumber
\end{align}
The Markov property for this system is expressed
as the dependence of $\mathbb{P}(X\!=\alpha_1 \alpha_2\ldots 
\alpha_L,t)$ only on its values at the immediately preceding time, 
$t-\delta t$ say. It is also assumed that the transition probabilities
are conditionally independent across different taxa.  This
is a standard assumption
\cite{SempleSteel2003,Felsenstein2004,SteelHendyPenny1998} and is 
quite
well founded from a biological perspective.  Again assuming
differentiability and linearity, a solution is found to be
\begin{align}
    \label{eq:multistoch}
P^{\alpha_1\alpha_2 \ldots\alpha_L}(t)=
&\sum_{\beta_1,\beta_2,\ldots,\beta_L}{M_1{}^{\alpha_{1}}}_{\beta_{1}}(t)
{M_2{}^{\alpha_{2}}}_{\beta_{2}}(t)\ldots 
{M_L{}^{\alpha_{L}}}_{\beta_{L}}(t)P^{\beta_1 \beta_2 \ldots 
\beta_L}(0).
\end{align}

The final part of the model is to introduce the branching. 
In the case of two taxa diverging from a common ancestor, 
considering that at the time of branching the character frequencies 
are identical, the correct formula for the pattern
frequencies is given by (see for example
\cite{RodriguezOliverMarinMedina1990})
\begin{align}
P^{\alpha_1\alpha_2 }(t)=&\sum_{\beta \in 
\mathcal{K}}{M_1{}^{\alpha_{1}}}_{\beta}(t) 
{M_2{}^{\alpha_{2}}}_{\beta}(t) p^{\beta}(0),
\end{align}
as will be derived in detail below. This situation can then be 
iterated for the case of arbitrary trees (see for example 
\cite{Farris1973,Felsenstein1981} as well as the standard texts 
already cited). 
Having given the standard stochastic model of phylogenetics we proceed
to abstract the presentation.  Introduce the vector space\footnote{
Although the above presentation involves only \textit{real}
numbers, we work over ${\mathbb C}$ to allow for the use of
more convenient symmetry adapted bases, or other ways of 
diagonalising rate matrices \cite{BashfordJarvisSumnerSteel2004} . Of 
course, measurable quantities are referred back to the distinguished 
basis at the end of the analysis.} 
$V\cong\mathbb{C}^K$, with preferred
basis $\{e_0,e_1,...,e_{K-1}\}$.  We associate the set of
probabilities (\ref{eq:singleprobs}) with the unique vector
\begin{align}
    & p^{\alpha}(t)\rightarrow p(t)=\sum_{\alpha\in
    {\mathcal K}}p^{\alpha}(t) e_\alpha.
\end{align}
Having made this abstraction it is possible to view the stochastic
evolution given by (\ref{eq:singlestoch}) as linear group action on
$V$, clearly an appropriate one parameter subgroup of $GL(K)$.  The
structure of the Markov group is discussed in \cite{Johnson1985}, and 
from
the viewpoint of invariant theory in \cite{SumnerJarvis2004,SumnerJarvis2005,SumnerBashfordJarvis2005}.
For the case of phylogenetics, the obvious generalisation is to 
the tensor product space $V^{\otimes L}$, 
with group action as the appropriate subgroup of the direct 
product group $GL(K)^{\times L}$.
The final step is to descibe the branching process upon this tensor 
product space.

In order to formalize this we introduce the \textit{splitting}
operator $\delta:V\rightarrow V\otimes V$.  Progress is made by simply
expressing the most general action of $\delta$ on the basis elements
of $V$:
\begin{align}\label{eq:splitting}
    \delta\cdot e_\alpha=&\sum_{\alpha, \beta, \gamma, 
}{\Gamma_\alpha}^{\beta 
    \gamma}e_\beta\otimes e_\gamma,
\end{align}
where ${\Gamma_\alpha}^{\beta \gamma}$ are an arbitrary set of 
coefficients.  
Imposing conditional independence upon the distinct 
branches in order to constrain these coefficients, we need
only consider initial probabilities of the form
\begin{align} 
    p_{(\lambda)}^{\alpha} &=\delta^\alpha_{\lambda},\quad \gamma 
    =0,1,...,K-1.
\end{align} 
Consider a branching even at time $t$ so that the initial single 
taxon 
state a small time before branching is
\begin{align}
    \label{eq:InitialP}
     p_{(\lambda)}(t) =\sum_{\alpha}p_{(\lambda)}^\alpha(t) e_\alpha 
&= 
     \sum_{\alpha}\delta_\lambda^\alpha 
     e_\alpha =  e_{\lambda}.                                  
\end{align}
Directly subsequent to the branching event the 2 taxon state is 
therefore given by
\begin{align}
     P_{(\lambda)}(t) &= \delta\cdot p_{(\lambda)}(t) 
     =\sum_{\alpha, \beta, 
     \gamma}\delta_\lambda^\sigma \Gamma_{\sigma}^{\rho \rho'} 
     e_{\rho}\otimes e_{\rho'}.
\end{align}
On the other hand, conditional independence 
leads to:
\begin{align}
    \label{eq:condindep}
     \mathbb{P}(X\!=\!\alpha_1 
     \alpha_2, & t+\delta t|X_1\!=\!X_2=\lambda, t) \nonumber\\
     =\mathbb{P}&(X_1\!=\!\alpha_1,t+\delta t|X_1\!=\!\lambda,t)
     \cdot 
     \mathbb{P}(X_2\!=\!\alpha_2,t+\delta t |X_2\!=\!\lambda,t).
\end{align}
Using the tensor formalism these transition probabilites can be 
expressed separately as
\begin{align}
&\mathbb{P}(X_1=\alpha_1,t+\delta 
t|X_1=\lambda,t)=\sum_{\rho}{{M_1}^{\alpha_1}}_{\rho}(\delta t)
p_{(\lambda)}^\rho(t),\nonumber\\
&\mathbb{P}(X_2=\alpha_2,t+\delta 
t|X_2=\lambda,t)=\sum_{\rho'}{{M_2}^{\alpha_2}}_{\rho'}(\delta t)
p_{(\lambda)}^{\rho'}(t). \end{align} However, from 
(\ref{eq:multistoch}) we have \begin{align}
&\mathbb{P}(X=\alpha_1\alpha_2,t+\delta 
t|X_1=X_2=\lambda,t)\nonumber\\
& \hspace{100pt}
=\sum_{\rho, \rho', \sigma} {{M_1}^{\alpha_{1}}}_{\rho}(\delta t) 
{{M_2}^{\alpha_{2}}}_{\rho'}(\delta t) \delta_\lambda^\sigma 
\Gamma_{\sigma}^{\rho \rho'}.\nonumber
\end{align}
Implementing (\ref{eq:condindep}), (\ref{eq:InitialP}) and 
considering the limit $\delta t 
\rightarrow 0$ with ${M^\alpha}_\rho(\delta t) \rightarrow {\delta^\alpha}_\rho$ then leads to the requirement that 
\begin{align}\label{eq:splitcond}
\Gamma_\lambda^{\rho \rho'}=&\,\delta_\lambda^\rho 
\delta_\lambda^{\rho'},
\end{align}
and the definition for the splitting operator in the preferred basis 
becomes simply
\begin{align}\label{eq:DeltaDefn}
\delta\cdot e_\alpha=&e_\alpha \otimes e_\alpha.
\end{align}

\begin{figure}[tbp]
		\centering
\raisebox{.5cm}{\begin{pspicture}(8,8)
\psset{framearc=0.0,framesep=0.0truecm,nodesep=8pt,unit=0.75truecm,
      arrowsize=4pt 2,arrowinset=0.4}
\def\pstlw{0.8pt}      
\psline{-}(0,0)(4,8)
\psline{-}(8,0)(4,8)
\psline{-}(4,8)(4,8.5)
\psline{-}(2.5,0)(5.25,5.5)
\psline{-}(4.25,3.5)(6,0)
\pscircle[linewidth=0.8pt,fillstyle=solid,fillcolor=white](4,8){.14} 
\pscircle[linewidth=0.8pt,fillstyle=solid,fillcolor=white](5.25,5.5){.14} 
\pscircle[linewidth=0.8pt,fillstyle=solid,fillcolor=white](4.25,3.5){.14} 
\rput(0,-.5){$\vec{1}$}
\rput(2.5,-.5){$\vec{2}$}
\rput(6,-.5){$\vec{4}$}
\rput(8,-.5){$\vec{8}$}
\rput(1.6,4.5){$M_{\vec{1}}$}
\rput(4.1,4.5){$M_{\vec{6}}$}
\rput(5.2,7){$M_{\vec{14}}$}
\rput(2.8,1.8){$M_{\vec{2}}$}
\rput(4.5,1.8){$M_{\vec{4}}$}
\rput(7.7,1.8){$M_{\vec{8}}$}
\rput(4,9){$p$}
\end{pspicture} }
\caption{The general Markov model for four taxa with tree 
$(1((23)4))$  (or $(\vec{1}((\vec{2}\vec{4})\vec{8}))$ in 
terms of binary edge labelling).  The $M$'s are arbitrary 
transition probabilities (Markov matrices) on the designated edges.}
	\label{fig:FourTreeFig}
\end{figure}
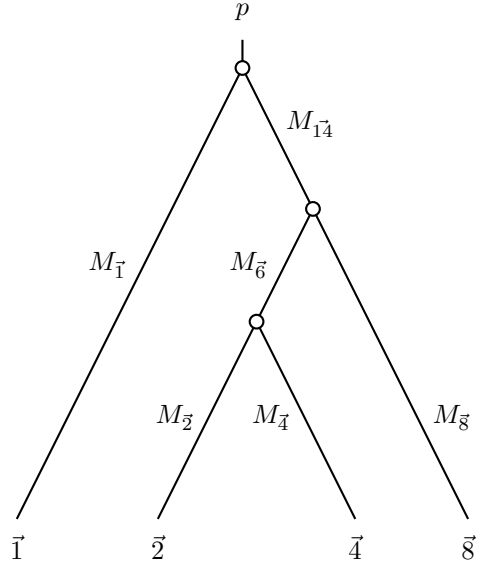

Using the above notation we are now in a position to write down 
formally expressions for 
the probabilities on arbitrary trees. As an example, the expression 
which defines the general
Markov model on the tree
$(1((23)4))$ (or\footnote{
In terms of the binary labelling introduced below, this tree is 
$(\vec{1}((\vec{2}\vec{4})\vec{8}))$, with the remaining edge 
assignments determined additively.} $(\vec{1}((\vec{2}\vec{4})\vec{8}))$) is given by (see figure \ref{fig:FourTreeFig})
\begin{align}
    \label{eq:DeltaEmbedding}
     P_{(\vec{1}((\vec{2}\vec{4})\vec{8}))}
&=(M_{\vec{1}}\!\otimes\!M_{\vec{2}}\!\otimes\!M_{\vec{4}}\!\otimes\!M_{\vec{8}})1\!\otimes\!\delta\!\otimes\!1
(1\!\otimes\!M_{\vec{6}}\!\otimes\!1)1\!\otimes\!\delta(1\!\otimes\!M_{\vec{14}})\delta\cdot 
     p
\end{align}
where $p$ is the initial single taxon distribution and the $M$'s 
are arbitrary stochastic operators (Markov matrices) on the designated edges. 
\section{Fock space and momentum labels for binary trees }
\label{sec:Fock}
In the previous section we have presented a description of
phylogenetic systems in terms of a multilinear tensor calculus based
on copies of the basic state space $V \simeq {\mathbb C}^{K}$.  This
comprises vectors with positive coefficients $p^{\alpha}$ in the
distinguished basis, corresponding to the theoretical probabilities
for observation of a particular character $\alpha$, $ \alpha =
0,\ldots,K-1$; higher rank tensors $P^{\alpha_{1}\alpha_{2}\ldots
\alpha_{n}}$ represent joint probability densities.  Moreover, we
introduced a linear operator $\delta: V \rightarrow V \otimes V$,
again defined by its matrix elements in the distinguished basis,
representing phylogenetic branching viewed dynamically as an event
occurring at a specific time.

In this and the following sections we wish to argue for a more
universal view which is appropriate for \textit{arbitrary} trees. 
Given that branching might occur at various times, this means that the
`state space' might be anything from $V$ (for the root edge of the
tree), to $V \otimes V$ (if there is only one branching), and so on,
up to $V \otimes V \otimes V \ldots \otimes V$, $L$ times, if the 
final
number of taxa (number of leaves of the tree) is $L$.  The only
logical way to encompass all these possibilities within one
description in linear algebra is to adopt as the proper state space,
an appropriate \textit{Fock space} ${ F}$ associated with $V$,
in this case for example
\begin{align}
\label{eq:FockDef}
{ F^{L}} = & {\mathbb C} \, \oplus \,  V \oplus \,  V \! \otimes \! 
                  V \,  \oplus 
                \cdots \oplus \, V \! \otimes \! V \otimes \! \cdots 
\otimes \! V 
                \nonumber \\
	     = &\displaystyle{\oplus}_{n=0}^{L} (\otimes^{n} V). 
\end{align}
The advantage of this formal change of perspective is that it allows 
both the normal time evolution (as described above, the Markov rate
operator acting on each copy of $V$), \textit{and} the branching 
operator (as described above, $\delta$, or its natural extensions 
$\one  \otimes \one \ldots \otimes \delta \otimes \ldots \one$, 
simultaneously to be regarded as operators on $ F$.

In physical settings it is conventional to apply the above 
construction for the description of `composite' systems 
where the state space $V$ corresponds to a single subsystem, and the 
tensor products allow for copies of $V$ corresponding to different 
numbers of subsystems. In relativistic systems this is of 
course the setting for elementary particle interactions, but the same 
idea is 
also appropriate in the nonrelativistic case. However, in quantum 
systems 
the Pauli principle mandates that the general spaces     
$V \! \otimes \! V \otimes \! \cdots \otimes \! V$ are too big -- 
the individual subsystems are \textit{indistinguishable} in that the 
ordering of individual state vectors in the tensor product is 
immaterial (up to a possible sign factor for fermionic systems). This 
means that the relevant Fock spaces are technically speaking the 
linear spaces ${ F}^{+}$, ${ F}^{-}$ associated respectively with the
symmetric (for bosons), or (for fermions) the antisymmetric or 
exterior tensor algebras:
\begin{align}
    \label{eq:FockDefFB}
    { F}^{+} = &\, {\mathbb C} \, \oplus \,  V \oplus \,  V \! \vee 
\! 
                  V \,  \oplus 
                \cdots \oplus \, V \! \vee \! V \vee \! \cdots \vee 
\! 
                V \cdots,
                \nonumber \\
		=& \oplus_{n=0}^{\infty} (\vee^{n} V); \nonumber \\
     { F}^{-} = & \, {\mathbb C} \, \oplus \,  V \oplus \,  V \! 
\wedge \! 
                  V \,  \oplus 
                \cdots \oplus \, V \! \wedge \! V \wedge \! \cdots 
\wedge \! 
                V,
                \nonumber \\
		=&\, \oplus_{n=0}^{L} (\wedge^{n} V)		
\end{align}

In adopting the machinery of Fock space to the phylogenetic context,
the `subsystems' become the individual taxa extant at any particular
stage of the branching process, and the (anti)symmetrisation principle
would need to be interpreted as saying that all taxa are equivalent, 
or that
the tensor probability density of rank $n$, is totally symmetric or
totally antisymmetric.  Thus for a given choice of observed characters
represented by the \textit{symmetric} probability density
$P^{\alpha_{1}\alpha_{2}\ldots\alpha_{n}}$, it would be immaterial 
which
taxon (from $1$ to $n$ in this case), carried which character:
\begin{align}
    P^{\alpha_{1}\alpha_{2}\ldots\alpha_{n}} = & \, P^{\alpha_{\sigma 
    1}\alpha_{\sigma 2}\ldots\alpha_{\sigma n}}
\end{align}
for any permutations $\sigma$. In phylogenetic branching, this symmetrisation
may well be appropriate for cases where it is suspected that a number 
of 
siblings are diverging from a common origin with \textit{equal} rate 
matrices\footnote{
A similar situation may apply in the antisymmetric case, but we shall 
not consider it further here.},
but in general, we would certainly wish to be able to 
distinguish between taxa.

These considerations imply that the higher rank tensor spaces $V \! 
\otimes \!  V \otimes \!  \cdots \otimes \!  V$ introduced above
should be regarded technically as products of a number of
\textit{labelled} spaces, for example for the final $L$ taxon system,
$V_{1} \!  \otimes \!  V_{2} \otimes \!  \cdots \otimes \!  V_{L}$
where each $V_{n}$ is a $\textit{distinct}$ copy of $V$, $V_{n} \simeq
V$, $n = 1,\ldots,L$.  However, since the $n$ taxon spaces required
for the system at earlier times (arising from branching at
intermediate nodes above the leaves of the tree) can comprise
\textit{any} subsets of the labels $1, \ldots, L$, we are led
necessarily to a labelling system apppropriate to the power set
$2^{L}$, or simply to the well-known system of edge labelling for
binary trees, by binary $L$-vectors, whereby leaf edges are labelled
by powers or decimal equivalents $1, 2^{1}, 2^{2}, \cdots
2^{L\!-\!1}$, and the assignments for the remaining edges determined
additively (for an example see figure \ref{fig:TreeFig}).

To this end we therefore introduce the following (extended) Fock 
space (we discuss only the bosonic case in this paper):
\begin{align}
    \label{eq:FockExtendedB}
    {\mathcal F}^{+} = &\, {\mathbb C} \, \oplus \,  {\mathcal V} 
\oplus \,  {\mathcal V} \! \vee \! 
                  {\mathcal V} \,  \oplus 
                \cdots \oplus \, {\mathcal V} \! \vee \! {\mathcal V} 
\vee \! \cdots \vee \! 
                {\mathcal V} + \cdots,
                \nonumber \\
		=&\,  \oplus_{n=0}^{\infty} (\vee^{n} {\mathcal V}); \nonumber \\
    {\mathcal V} :=& \sum_{{ \mathbf k} \in \pi 
    {{\mathbb Z}_{2}}^{L}} \oplus V_{{\mathbf k}}.
\end{align}   
The linear operators which can be used to construct the branching 
operator are defined in terms of the so-called creation 
and annihilation operators on ${\mathcal F}^{+}$. For 
$v_{{\mathbf k}} \in 
V_{{\mathbf k}}$, $v^{{\mathbf k}*} \in 
V_{{\mathbf k}}^{*}$ define the operators 
$a^{\dag}(v_{{\mathbf k}} ): \vee^{n} {\mathcal V} 
\rightarrow \vee^{n+1} {\mathcal V}$,
$a(v^{{\mathbf k}*}):  \vee^{n} {\mathcal V} \rightarrow
\vee^{n-1} {\mathcal V}$ by (see for example \cite{Sudbery1986})
\begin{align}
a^{\dagger}(v_{ {\mathbf k}} )\cdot v_{{\mathbf k}_{1}} 
\vee 
         v_{{\mathbf k}_{2} }\vee \cdots \vee v_{ {\mathbf k}_{n}}
        = & v_{ {\mathbf k}} \vee  v_{{\mathbf k}_{1}} 
\vee v_{{\mathbf k}_{2}} 
        \vee \ldots \vee v_{{\mathbf k}_{n}}; 
       \nonumber \\
     a(v^{ {\mathbf k} *}) \cdot v_{{\mathbf k}_{1}} 
\vee 
     v_{{\mathbf k}_{2}} \vee \ldots \vee v_{{\mathbf k}_{n} } 
     = &  
      \sum_{m=1}^{n} 
      {\delta^{{\mathbf k}}}_{{\mathbf k}_{m}} 
v^{{\mathbf k}*}( v_{ {\mathbf k}_{m} } )
      v_{ {\mathbf k}_{1} }\vee \ldots \widehat{v}_{ {\mathbf k }_{m} } 
      \ldots \vee v_{ {\mathbf k}_{n} } \nonumber      
\end{align}
where $\widehat{v}$ denotes the omission of the corresponding vector
(the dual action has been formally extended to be zero on differently 
labelled spaces, and the corresponding ${\delta^{ \mathbf k}}_{{\mathbf k}_{m}}$ factor displayed explicitly). 
The operators so defined should then be formally summed to give
operators on the whole of ${\mathcal F}^{+}$ (and by definition $a^{
{\mathbf k} }(v^{*})\cdot {\mathbb C} =0$), for which we retain
the same symbol.  In particular for the unit vectors
$e_{{\mathbf k}\alpha}$ and their duals $e^{{\mathbf k}\alpha}$ we define
\begin{align}
    \label{eq:DefnAAdagger}
    a^{\dagger}(e_{{\mathbf k}\alpha}) := & 
    a^{\dagger}_{{\mathbf k}\alpha}, \qquad  
    a(e^{{\mathbf k}\alpha}) :=  a^{{\mathbf k}\alpha}.
\end{align}
The operators $a(u^{*})$, $a^{\dag}(v)$ fulfil the commutation 
(ordering) relations
$a(u^{*}) a^{\dag}(v) - a^{\dag}(v) a(u^{*}) \equiv {[} a(u^{*}), 
a^{\dag}(v) {]} = u^{*}(v) \one$, where $\one$ is the unit operator 
on ${\mathcal F}^{+}$; 
in particular for the mode operators 
$a^{\dagger}_{{\mathbf k}\alpha}$, $a^{{\mathbf k}\alpha}$ we have
\begin{align}
    \label{eq:CommRelns}
    {[}a^{{\mathbf k}\alpha}, a^{\dagger}_{{\mathbf l} 
\beta}{]} 
    = & {\delta^{{\mathbf k}} }_{{\mathbf l}} 
{\delta^{\alpha} }_{\beta} \one. 
\end{align}
Moreover if we define the `ground' state to be $\one \in {\mathbb 
C}$, we have the algebraic means to write an \textit{arbitrary} 
element of the corresponding distinguished basis in Fock space,
\begin{align}
    e_{{\mathbf k}_{1}\alpha_{1}} \vee e_{{\mathbf k}_{2}\alpha_{2}} 
    \vee \cdots e_{{\mathbf k}_{n}\alpha_{n}}
    := & a^{\dagger}_{{\mathbf k}_{1} \alpha_{1}}\cdot \mbox{}
    a^{\dagger}_{{\mathbf k}_{2} \alpha_{2}}\cdots
    a^{\dagger}_{{\mathbf k}_{n} \alpha_{n}} \cdot \one .
\end{align}
In what follows it will be notationally more compact to introduce the 
so-called Dirac bra-ket notation for vectors in ${\mathcal V}$ and 
their duals. Thus formally we write
\begin{align}
    \one \leftrightarrow & | 0 \rangle, \qquad \one^{*} 
    \leftrightarrow \langle 0 |; \nonumber \\
e_{{\mathbf k}\alpha} \leftrightarrow & 
a^{\dagger}_{{\mathbf k} \alpha }| 0 \rangle 
 \equiv |{\mathbf k},\alpha  \rangle, \qquad  e^{* 
{\mathbf k}\alpha}  
\leftrightarrow \langle 0 | a^{{\mathbf k} \alpha} 
\equiv \langle {\mathbf k},\alpha |; \nonumber \\
e_{{\mathbf k}_{1}\alpha_{1}} \vee e_{{\mathbf k}_{2}\alpha_{2}} 
\vee \cdots e_{{\mathbf k}_{n}\alpha_{n}}
:= & a^{\dagger}_{{\mathbf k}_{1} \alpha_{1}}\cdot \mbox{}
a^{\dagger}_{{\mathbf k}_{2} \alpha_{2}}\cdots
a^{\dagger}_{{\mathbf k}_{n} \alpha_{n}} |0 \rangle 
    \equiv |{\mathbf k}_{1} \alpha_{1}, {\mathbf k}_{2} 
    \alpha_{2}, \cdots {\mathbf k}_{n} \alpha_{n} \rangle 
    \nonumber, 
\end{align}
where the latter list may include repetition. In this case the 
explicit notation 
\begin{align}
    \label{eq:PcreationDefn}
    |{\mathbf k}_{1} \alpha_{1},m_{1}; {\mathbf k}_{2} 
    \alpha_{2}, m_{2}; \cdots {\mathbf k}_{r} \alpha_{r},m_{r} 
\rangle 
    = & (a^{\dagger}_{{\mathbf k}_{1} \alpha_{1}})^{m_{1}} 
\cdot
    (a^{\dagger}_{{\mathbf k}_{2} \alpha_{2}})^{m_{2}} \cdots 
    (a^{\dagger}_{{\mathbf k}_{r} \alpha_{r}})^{m_{r}} | 0 
    \rangle , 
\end{align}
(corresponding to the so-called \textit{number basis}) is occasionally
mandatory.  Finally we introduce the natural Cartesian inner product
on these state vectors (with the $e_{{\mathbf k}\alpha}$
orthonormal in ${\mathcal V}$), extended to ${\mathcal F}^{+}$ in such
a way that each creation and annihilation pair is mutually hermitean,
and in general
\begin{align}
    \langle {\mathbf k}_{1} \alpha_{1},m_{1}; {\mathbf k}_{2} 
    \alpha_{2}, m_{2}; \cdots {\mathbf k}_{r} \alpha_{r},m_{r} 
    |{\mathbf l}_{1} \beta_{1},n_{1}; {\mathbf l}_{2} 
    \beta_{2}, n_{2}; \cdots {\mathbf l}_{s} \beta_{s},n_{s} 
\rangle
    =& \delta_{rs} \prod_{q=1}^{r} \delta_{{\mathbf k}_{q} 
    {\mathbf l}_{q}} 
    \delta_{\alpha_{q}\beta_{q}} \cdot \delta_{m_{q} n_{q}}m_{q}!
\end{align}
Although the general structure will be needed in the formalism below, 
sample states are in practice those belonging to 
a fixed number $n$ of subsystems (for example $n = L$, the number of 
taxa), with (distinct) labelled momenta without multiplicity, of the 
general form
\begin{align}
    \label{eq:TensorState}
    | P \rangle & = \sum_{\alpha_{1}, \alpha_{2}, \cdots \alpha_{n} 
    =0}^{K-1} P^{\alpha_{1}\alpha_{2} \cdots 
\alpha_{n}}|{\mathbf k}_{1} \alpha_{1}, {\mathbf k}_{2} 
    \alpha_{2}, \cdots {\mathbf k}_{n} \alpha_{n} \rangle .
\end{align}
Such state vectors can immediately be attributed to a theoretical 
probability 
density for $n$ taxa provided that the coefficients are positive and 
that their sum is unity. For technical reasons we introduce an 
auxiliary `reservoir' state (dual, or `bra' vector)
\begin{align}
    \mbox{}^{(n)}\langle \Omega | = & \sum_{\alpha_{1}, \alpha_{2}, 
\cdots \alpha_{n} 
    =0}^{K-1} \langle {\mathbf k}_{1} \alpha_{1}, 
{\mathbf k}_{2} 
    \alpha_{2}, \cdots {\mathbf k}_{n} \alpha_{n} | \nonumber
\end{align}
so that this condition can be written
\begin{align}
    \mbox{}^{(n)}\langle \Omega |P \rangle = 1 & \qquad 
\longleftrightarrow 
    \, \sum_{\alpha_{1}, \alpha_{2}, \cdots \alpha_{n} 
    =0}^{K-1} P^{\alpha_{1}\alpha_{2} \cdots \alpha_{n}} = 1.
\end{align}
In full generality, the auxiliary vector (allowing for multiplicities
and summing over different momenta) becomes
\begin{align}
    \langle \Omega | := & \langle 0 | e^{\sum_{ {\mathbf k} 
\in \pi {\mathbb Z}_{2}^{L} } 
    \sum_{\alpha=0}^{K-1} a^{{\mathbf k} \alpha}}, \quad 
    \mbox{or} \quad \\
    \mbox{}^{\chi}\langle \Omega | = & \langle 0 | e^{ 
\sum_{{\mathbf k} \in \pi {\mathbb Z}_{2}^{L} }  
    \sum_{\alpha=0}^{K-1} \chi_{ {\mathbf k} \alpha }  
a^{{\mathbf k} \alpha}   },
\end{align}
where the latter form is convenient for notational purposes (with the 
understanding that $\chi_{ {\mathbf k} \alpha } \rightarrow 
1$).

We shall be concerned with general functions $f$ (such as the
probabilities $P$, and below with operators built from the creation
and annihilation mode operators) which are obtained as formal sums of
terms depending on the `momentum' labels, say $f_{{\mathbf k}\! 
\!  {\mathbf l}\cdots}$.  With the convention we have adopted
(of scaling the $\mathbf k$'s by $\pi$) to any such function we can
associate functions over a dual space ${\mathbf x},
{\mathbf y} \cdots \in {\mathbb Z}_{2}^{L}$ by a formal Fourier
transform.  This is of course the discrete Fourier-Hadamard
transformation (the phase factors are simply $\pm 1$), and the
functions $f$ on `configuration' (position) space are periodic with
periods $2 {\mathbf a}$ for ${\mathbf a} \in {\mathbb
Z}_{2}^{L}$.  In particular for the constant function in one variable 
$\one_{{\mathbf k}}=1$,
\[
    \delta({\mathbf x}) = \left(\frac{1}{2^{L}}\right)  
\sum_{{\mathbf k} \in 
    \pi{\mathbb Z}_{2}^{L}} e^{i({\mathbf k} \cdot 
{\mathbf x})}, 
    \qquad 
    1 = \sum_{{\mathbf x} \in {\mathbb Z}_{2}^{L}} 
    \delta({\mathbf x}) e^{-i({\mathbf k} \cdot 
{\mathbf xs})},
\]
(where $\delta({\mathbf x}) = \delta({\mathbf x},{\mathbf 0}))$.  More generally,
\begin{align}
    f({\mathbf x}) := & \left( 
    \frac{1}{2^{L}}\right) \sum_{{\mathbf k} \in 
    \pi{\mathbb Z}_{2}^{L}} 
    f_{{\mathbf k}} 
    e^{i({\mathbf k} \cdot {\mathbf x})}, \qquad
    f_{{\mathbf k}} = \sum_{{\mathbf x} \in 
    {\mathbb Z}_{2}^{L}} f({\mathbf x}) e^{-i({\mathbf k} \cdot {\mathbf x})}.
\end{align}
In two variables, we have in turn
\begin{align}
    f({\mathbf x}, {\mathbf y}) := & \left( 
    \frac{1}{2^{L}}\right) ^{2} \sum_{{\mathbf k} \in 
    \pi{\mathbb Z}_{2}^{L}} \sum_{{\mathbf l} \in \pi{\mathbb 
Z}_{2}^{L}} 
    f_{{\mathbf k} {\mathbf l}} 
    e^{i({\mathbf k} \cdot {\mathbf x} + {\mathbf l} 
    \cdot {\mathbf y})}, \qquad 
    f_{{\mathbf k} {\mathbf l}} = \sum_{{\mathbf x} \in 
    {\mathbb Z}_{2}^{L}} f({\mathbf x}, {\mathbf y}) 
e^{-i({\mathbf k} \cdot {\mathbf x} + 
    {\mathbf l} \cdot {\mathbf y})},
\end{align}
and generally 
\[
f({\mathbf x}+ 2{\mathbf a}, {\mathbf y}+2{\mathbf b}, \ldots )
 = f({\mathbf x}, {\mathbf y}, \ldots).
\]

As an example of the creation and annihilator formalism, let us give 
an operator on ${\mathcal F}^{+}$ equivalent to the branching 
operator $\delta: V \rightarrow V \otimes V$ introduced above (which 
has to be extended case by case to allow for branchings on particular 
factors of $\otimes^{n}V$ for a particular tree). Recall that the 
general form 
\begin{align}
    \label{eq:GammaMatrixForm}
\delta (e_{\alpha}) = &  {\Gamma_{\alpha}}^{\beta \gamma} e_{\beta} 
\otimes e_{\gamma}
\end{align}
was subsequently specialised to ${\Gamma_{\alpha}}^{\beta \gamma} = 
{\delta^{\alpha}}_{\beta} {\delta^{\alpha}}_{\gamma}$ on the basis of 
conditional independence. Next assume that the copies of $V$ 
involved are distinguished by different labels  ${\mathbf k}, 
{\mathbf l}, {\mathbf m}$ so that there is no 
difference between the above use of $\otimes$ and the correct $\vee$ 
as far as the symmetric algebra is concerned (below we shall see that 
the momentum labels are such that ${\mathbf k} = {\mathbf l} \!+\! {\mathbf m}$). Consider then the 
operator ${\Delta} = \sum_{\alpha, \beta, \gamma} 
{\Gamma_{\alpha}}^{\beta \gamma} a^{\alpha}
 {a^{\dagger}}_{\beta} {a^{\dagger}}_{\gamma}$, and its action on a 
state 
 $V \ni |p \rangle = p^{0}|0\rangle + p^{1}|1\rangle
\ldots + p^{K-1} |K\! - \! 1 \rangle =  \sum_{\xi}p^{\xi}|\xi 
\rangle$ :
\begin{align}
    \label{eq:GammaTrilinear}
  {\Delta} |p \rangle =  & \sum_{\alpha, \beta, \gamma} 
{\Gamma_{\alpha}}^{\beta \gamma}
  {a^{\dagger}}_{\beta} {a^{\dagger}}_{\gamma}  a^{\alpha} 
\sum_{\xi}p^{\xi}{a^{\dagger}}_{\xi}|0 \rangle 
  \\
   = &   \sum_{\alpha, \beta, \gamma} {\Gamma_{\alpha}}^{\beta \gamma}
  {a^{\dagger}}_{\beta} {a^{\dagger}}_{\gamma} \sum_{\xi} 
  p^{\xi}{[}a^{\alpha},{a^{\dagger}}_{\xi}{[}|0 \rangle \nonumber \\
  =&  \sum_{\alpha, \beta, \gamma} {\Gamma_{\alpha}}^{\beta \gamma}
  {a^{\dagger}}_{\beta} {a^{\dagger}}_{\gamma}  
  p^{\alpha}|0 \rangle \nonumber \\
  =& \sum_{\alpha, \beta, \gamma} 
{[}p^{\alpha}{\Gamma_{\alpha}}^{\beta \gamma}{]}
   |\beta , \gamma \rangle. \nonumber
\end{align}
Thus, indeed, the requisite branching from the initial ancestral 
density $|p \rangle = \sum_{\alpha} p^{\alpha} |\alpha \rangle$ to 
the density for 2 taxa after branching, with characters shared 
equally ($|P \rangle = \sum_{\alpha} p^{\alpha} |\alpha, \alpha 
\rangle $ for the special choice (\ref{eq:splitcond}) of $\Gamma$), 
has 
been effected, and the operator $\Delta$ provides a generalisation of 
the splitting operator $\delta$ of (\ref{eq:splitting}), 
(\ref{eq:DeltaDefn}) suitable for 
representing embeddings of the latter on individual factors of the 
tensor product, as in (\ref{eq:DeltaEmbedding}).
In \S \ref{sec:Interactions} below, we return to this operator in the 
context of a dynamical change model for branching. As will 
be seen, it needs to be embellished by edge `momentum'  
labels in order to generate appropriate phylogenetic trees, and also 
to be assigned a time dependence corresponding to the fact that 
branching events will occur at specific times in an evolutionary 
sense. 
These apparent complications need to be contrasted with the 
fact that if the splitting operator $\delta$ is used in its original 
form, for a specific tree, its action on tensor products must be 
extended on a case-by-case basis, as in (\ref{eq:DeltaEmbedding}). 
In \S \S \ref{sec:PathIntRev}, \ref{sec:FreePathInt} below, we turn 
to 
a review of the path integral method for solving the time evolution 
of systems described in the operator language, and then apply the 
technique to a system of taxa which is `free', that is, 
evolving without any phylogenetic association, after having developed 
the appropriate form for the rate operator of such a system.

\section{Review of path integral formalism}
\label{sec:PathIntRev}

In this section we review briefly the path integral formalism for the
representation of the time development of stochastic processes whose
`microscopic' states represent probabilities of certain `particle'
numbers at each time.  The aim of the next section will be to apply
the technique to the multi-linear representation of taxonomic states
developed in \S \ref{sec:TensorMethods} and transcribed into the
`occupation number' representation in \S \ref{sec:Fock} above.  The
task at hand is to transcribe the abstract occupation number
representation (as developed for our purposes in the previous section)
into a formalism of integral operators acting on generating functions
representing the appropriate probability densities.  This section
closely follows the presentation of Peliti\cite{Peliti1985}.

For a single system we therefore have microscopic states of the form
(see (\ref{eq:DefnAAdagger})) $|n \rangle = a^{\dagger}\mbox{}^{n}| 
0 \rangle$, with the creation and annihilation operators 
$a^{\dagger}$ 
and $a$  being hermitean conjugates of each other with
\begin{align}
a^{\dagger}| n \rangle =& | n+1 \rangle, \qquad a| n \rangle = | n-1 
\rangle, \nonumber \\
\langle n | m \rangle =& n! \delta_{mn}. 
    \label{eq:ScalarProd}
\end{align}
Next we make the transcription from states  
\[
|\phi \rangle = \sum_{n=0}^{\infty} \phi^{n} |n \rangle, \qquad 
\langle \phi | \psi \rangle = \sum_{n=0}^{\infty} n! \phi_{n}\psi_{n}
\]
to a space of functions
\[
|\phi \rangle \leftrightarrow \phi(z) = \sum_{n=0}^{\infty} \phi_{n} 
z^{n},
\]
where the variable `z' is a formal variable if (as in the usual
statistical context) $\phi(z)$ is meant as a formal generating
function.  However for the present development it is convenient to 
allow
$z$ to be complex and to regard the $\phi(z)$ as analytic functions 
belonging to a Hilbert space. 
In terms of defining path integrals $z$ can be taken to be
real, or analytically continued subject to certain
prescribed asymptotic behaviour (possibly together with
constraints forcing its passage through specified points of the 
complex
plane). 

Using the elementary identity
\begin{align}
    \label{eq:deltaidentity}
    n! \delta_{mn} =& \int dz z^{n}(-\frac{d}{dz})^{m} \delta(z)
\end{align}
(which can be established by integration by parts) the scalar 
product (\ref{eq:ScalarProd}) becomes
\begin{align}
    \label{eq:ScalarProdInt}
 \langle \phi | \psi \rangle = & \int dz \phi({z}) 
\psi(-\frac{d}{dz}) 
 \delta(z), \qquad \mbox{or} \nonumber \\
\langle \phi | \psi \rangle = & \int \frac{d{z} d\zeta}{2 \pi } 
\phi({z}) \psi(i\zeta) e^{-i{z} \zeta }.  
\end{align}
Associated with the matrix elements ${\mathcal A}_{mn} = \langle m | 
A | n 
\rangle$ of any operator in the number 
basis is the integral kernel ${\mathcal A}(\bar{z}, \zeta)$ 
\begin{align}
{\mathcal A}(z, \zeta) =& \sum_{m,n=0}^{\infty} 
\frac{{{z}}^{m}}{m!} A_{mn} \frac{\zeta^{n}}{n!}
\end{align}
such that 
\begin{align}
    |\psi \rangle =& A |\phi \rangle = \sum_{m,n} \frac{|m \rangle 
    \langle m |}{m!} A \frac {|n \rangle 
    \langle n |}{n!} | \phi \rangle \nonumber 
\end{align}
can be expressed via $\psi(z) = \sum \psi_{m}z^{m}$, with
\begin{align}
   \psi(z) = & \int 
    \frac{d{\zeta} d\zeta'}{2 \pi }
    {\mathcal A}(z,\zeta)\phi(i\zeta') e^{-i \zeta \zeta'} 
    \label{eq:OpAction}
\end{align}
easily established using the identity (\ref{eq:deltaidentity}) above.
Similarly the integral kernels of the product ${\mathcal A}{\mathcal 
B}$ of two operators ${\mathcal A}$, 
${\mathcal B}$ is:
\begin{align}
    \label{eq:OperatorProduct}
    {\mathcal A}{\mathcal B}(z,\zeta) =& \int \frac{d{\eta} d\eta'}{2 
\pi }{\mathcal A}(z,\eta) 
    {\mathcal B}(i\eta', \zeta)  e^{-i \eta \eta'} 
\end{align}
The integral kernel ${\mathcal A}(z,\zeta)$ has a natural 
combinatorial connection to the normal kernel
${A}(z,\zeta)$ where $A$ is expressed in terms of creation 
and annihilation operators:
\begin{align}
    A = & \sum_{m,n =0}^{\infty} {a^{\dagger}}\mbox{}^{m}{ 
    A}_{mn} a^{n}, \quad \mbox{define} \nonumber\\
    {A}(z, \zeta) := & \sum_{m,n} {z}^{m}{ 
    A}_{mn} \zeta^{n}. \nonumber 
\end{align}
Then there is the simple relationship
\begin{align}
    {\mathcal A}(z,\zeta) =& e^{z \zeta} {A}(z, \zeta). \nonumber 
\end{align}

Consider the effect of stochastic time 
evolution on the system. In the linear case the state 
probabilities are assumed to change according to the master equation
\begin{align}
    \label{eq:ClassicalMarkov}
    \frac{d}{dt}\phi_{n} =& \sum_{m \ne n} \left( r_{m \rightarrow n} 
    \phi_{m} - r_{n \rightarrow m}\phi_{n} \right)
\end{align}
where $r_{m \rightarrow n}$ are the transition rates. 
It is convenient to define 
$R_{nm} = r_{m \rightarrow n}$ and $R_{nn} = - \sum_{m} r_{n 
\rightarrow 
m} = - \sum_{m}R_{mn} $ so that the time evolution becomes
\begin{align}
    \label{eq:LinearMarkov}
    \frac{d}{dt}\phi_{n} =& \sum_{m}R_{nm} \phi_{m}
\end{align}
with the understanding that the rate matrix satisfies $\sum_{n} 
R_{nm} =0$, for all $m$, or introducing the reservoir state $\langle 
\Omega |$ from above, and 
regarding $R(t)$ as an operator on state space which can be time 
dependent\footnote{
This entails $d{\phi_{n}}/dt = \langle n | R | \phi \rangle / n!$, 
consistent with the resolution of the identity (see 
(\ref{eq:ScalarProdInt}) above).}
\begin{align}
    \frac{d}{dt}| \phi(t) \rangle = & R(t) |\phi(t) \rangle, \quad 
    \mbox{with} \quad 
    \langle \Omega | R(t) = 0.
\end{align}
   
With the above notation we can now develop a path integral 
representation for the evolution kernel of the system.     
Approximate the form of evolution operator for a small change as 
$M_{(t+\delta t, t)} 
\simeq e^{R(t)\delta t}$, and for the evolution operator as a whole 
as 
a product of infinitesimal changes
\begin{align}
    M_{(T,0)} \simeq & M_{(T, T-\delta t)} \cdot M_{(T - \delta t, T 
- 2\delta 
    t)} \cdots M_{(2 \delta t, \delta t)} \cdot M_{(\delta t, 0)}. 
\nonumber
\end{align}
Approximating each of the exponentials by a linear expression, and 
using the above relation between normal and integral kernels leads to
\begin{align}
    {\mathcal M}_{(t+\delta t, t)}(z, \zeta) \simeq & e^{z 
    \zeta}(1+\delta t R(t)(z,\zeta)). \nonumber
\end{align}
Using this and iterating (\ref{eq:OperatorProduct})to give a multiple 
integral 
representation of this product, assuming $T = N \delta t$, we have
\begin{align}
    M_{(T,0)}(z, \zeta) \simeq & \int M_{(T, T-\delta t)}(z, 
\eta_{1}) 
    \frac{d\eta_{1}d\eta_{1}'e^{-i\eta_{1}\eta_{1}'}}{2 \pi}M_{(T - 
\delta t, T - 2\delta 
    t)}(\eta_{1}', \eta_{2}) 
\frac{d\eta_{2}d\eta_{2}'e^{-i\eta_{2}\eta_{2}'}}{2 \pi} 
    \cdots \cdot \nonumber \\
    & \cdot M_{(2 \delta t, \delta t)}(\eta_{N-2}', \eta_{N-1})  
    \frac{d\eta_{N-1}d\eta_{N-1}'e^{-i\eta_{N-1}\eta_{N-1}'}}{2 \pi}  
M_{(\delta t, 
    0)}(\eta_{N-1}', \zeta) \nonumber \\
    \simeq & 
    \int \prod_{\ell = 0}^{N-1} 
    \frac{d\eta_{\ell}d\eta_{\ell}'}{2 \pi}
    \cdot \exp{\left(\sum_{\ell=0}^{N-1}\left[ 
-i\eta_{\ell+1}'(\eta_{\ell+1} - \eta_{\ell}) + \delta t 
    R(t)(i\eta_{\ell+1}', \eta_{\ell}) \right]\right)} \cdot 
    e^{z\eta_{N}} 
\end{align}
which leads formally in the limit $N \rightarrow \infty$ to the path 
integral 
representation (\textit{c.f.} \cite{Peliti1985} equations (2.23), 
(2.24))
\begin{align}
    \label{eq:PathIntRep}
    {\mathcal M}_{T}(z, \zeta) = & \int d[\eta]d[\eta'] 
    \exp{\left( \int_{0}^{T} dt \left( -i 
\eta'(t)\stackrel{\scriptscriptstyle{\bullet}}{\eta}(t) + iR_{t}(i 
\eta', \eta) \right) 
    +{z \eta(T)} \right)}.
\end{align}    
Here the $2 \pi$ factors have been incorporated into the path 
integral measure, and the integrations over paths $\eta(t)$, 
$\eta'(t)$ from $0$ to $T$ are made with the boundary conditions on 
each endpoint given by 
\begin{align}
    \label{eq:BoundaryConditions}
    \eta(0) = \zeta, \quad & i\eta'(T) = z.
\end{align}
The additional boundary term $\exp{(z \eta(T))}$ also arises from the 
continuum ($N \rightarrow \infty$) limit of the iterated product 
representation.

It is important to point out that the path integral representation 
\cite{Peliti1985,Doi1976a,Doi1976b} 
also allows closed form expressions to be written down for the means 
(and in 
principle higher moments) of any desired observable quantities. This 
has not only formal significance but also, depending on the operator, 
opens an avenue for explicit analytical calculations. 
\section{Evolution kernel for free phylogenetic system}
\label{sec:FreePathInt}

With our review of path integrals for stochastic systems in hand, 
we now return to the discussion of phylogenetic systems in the 
notation of \S \ref{sec:Fock}. We concentrate here on the `free' 
system, 
that is, phylogenetic evolution without phylogenetic branching. As we 
now argue, the normal kernel of the rate operator can be taken to be 
quadratic, so that the entire path integral assumes Gaussian form and 
admits 
a formal steepest descent evaluation. In the next section we also 
introduce 
interactions along the lines of (\ref{eq:GammaTrilinear}) and 
indicate 
in simple examples which indeed reproduce the expected evolution (at 
least if the rate matrix is time independent) that this leads to the 
correct 
probabilities\footnote{
The formalism also applies to the inhomogeneous case (time-dependent 
rates), provided that the `propagator' is known (see below).}.

Transition rates have been discussed in \S \ref{sec:TensorMethods} in 
the tensor formalism, 
and in \S \ref{sec:PathIntRev} in introducing the path integral 
representation. However in 
the context of \S \ref{sec:Fock}, the appropriate time evolution must 
be the assignment of a rate operator to each possible edge `momentum' 
label, ${\mathbf k} \in \pi{\mathbb Z}_{2}^{L}$. In contrast to 
(\ref{eq:ClassicalMarkov}) then, in which it is assumed that the 
rates 
$r_{m\rightarrow n}$ and the $|m \rangle $, $|n \rangle $ refer to 
\textit{differing} occupation numbers, we thus construct initially a 
\textit{number-conserving} rate operator, at least inasmuch as the
`particle number' operator does not distinguish between the character 
types $\alpha = 0, 1, \ldots, K\!-\!1$ which take on the status of 
`internal' degrees of freedom. Indeed the number operator for 
edge ${\mathbf k}$ is 
\begin{align}
    N_{{\mathbf k}} = & 
    \sum_{\alpha} a^{\dagger}_{{\mathbf k}\alpha} 
    a^{{\mathbf k}\alpha}, \quad \mbox{such that} \nonumber \\
    {[}N_{{\mathbf k}},  a^{\dagger}_{{\mathbf k}\alpha} {]}  = 
a^{\dagger}_{{\mathbf k}\alpha}, \qquad & {[}N_{{\mathbf k}},  
a^{{\mathbf k}\alpha} {]}
    = - a^{{\mathbf k}\alpha}, \nonumber \\
    {[}N_{{\mathbf k}},  a^{\dagger}_{{\mathbf k}\alpha} 
    a^{{\mathbf k}\beta}{]} =& 0 \quad \forall \, \alpha, \beta
\end{align}
This means of course that the rate operator must be bilinear in both 
creation and annihilation operators of type ${\mathbf k}$, leading to 
the second-quantised expression
\begin{align}
    R = \sum_{{\mathbf k} \in \pi{\mathbb Z}_{2}^{L}} 
R_{{\mathbf k}} =& 
    \sum_{{\mathbf k}\in \pi{\mathbb Z}_{2}^{L}}\sum_{\alpha, 
\beta} a^{\dagger}_{{\mathbf k}\alpha}
    {{R_{{\mathbf k}}}^{\alpha}}_{\beta} a^{{\mathbf k}\beta}.
    \label{eq:FreeRateOpMomSpace}
\end{align}
As mentioned in \S \ref{sec:Fock} above, we will be 
concerned with \textit{single} occupation numbers for each momentum 
mode (for generalisations see the concluding remarks in 
\S \ref{sec:Concl} below).
Thus for a general tensor state (\ref{eq:TensorState}) (\textit{c.f.} 
\S 
\ref{sec:TensorMethods}), 
\begin{align}
   | \stackrel{\scriptscriptstyle{\bullet}}{P} \rangle & = R | {P} 
\rangle \nonumber \\
   & = \sum_{\gamma_{1}, \gamma_{2}, \cdots \gamma_{n} 
    =0}^{K-1} P^{\gamma_{1}\gamma_{2} \cdots \gamma_{n}}
    R|{\mathbf k}_{1} \gamma_{1}, {\mathbf k}_{2} 
    \gamma_{2}, \cdots {\mathbf k}_{n} \gamma_{n} \rangle. 
    \nonumber 
\end{align}
Using the fundamental relation
\begin{align}
    {[} R, a^{\dagger}_{{\mathbf l}\gamma} {]} & =
    {[}\sum_{{\mathbf k}\in \pi{\mathbb 
Z}_{2}^{L}}\sum_{\alpha, \beta} a^{\dagger}_{{\mathbf k}\alpha}
    {{R_{{\mathbf k}}}^{\alpha}}_{\beta} a^{{\mathbf k}\beta}, a^{\dagger}_{{\mathbf l}\gamma} {]} \nonumber \\
    & = \sum_{{\mathbf k}\in \pi{\mathbb 
Z}_{2}^{L}}\sum_{\alpha, \beta} 
   a^{\dagger}_{{\mathbf k}\alpha}
    {{R_{{\mathbf k}}}^{\alpha}}_{\beta}  {[} 
a^{{\mathbf k}\beta}, a^{\dagger}_{{\mathbf l}\gamma} {]} \nonumber \\
    & = \sum_{\alpha} 
   a^{\dagger}_{{\mathbf l}\alpha}
    {{R_{{\mathbf l}}}^{\alpha}}_{\gamma},
     \label{Rcommutator}
\end{align}
where (\ref{eq:CommRelns}) has been used, 
together with (\ref{eq:PcreationDefn}), we find finally as 
required
\begin{align}
    \label{eq:PtensorEvolution}
     | \stackrel{\scriptscriptstyle{\bullet}}{P} \rangle 
   & \equiv \sum_{\gamma_{1}, \gamma_{2}, \cdots \gamma_{n} 
    =0}^{K-1} 
\stackrel{\scriptscriptstyle{\bullet}}{P}^{\gamma_{1}\gamma_{2} 
\cdots \gamma_{n}}
    |{\mathbf k}_{1} \gamma_{1}, {\mathbf k}_{2} 
    \gamma_{2}, \cdots {\mathbf k}_{n} \gamma_{n} \rangle, 
    \quad \mbox{where} \nonumber \\ 
    \stackrel{\scriptscriptstyle{\bullet}}{P}^{\gamma_{1}\gamma_{2} 
\cdots \gamma_{n}} & =
    \sum_{\gamma} \left( {{R_{{\mathbf  k_{1}}}}^{\gamma_{1}}}_{\gamma}
    {P}^{\gamma \gamma_{2} \cdots \gamma_{n}} + {{R_{{\mathbf k_{2}}}}^{\gamma_{2}}}_{\gamma}
    {P}^{\gamma_{1} \gamma \cdots \gamma_{n}} + \cdots 
    {{R_{{\mathbf k_{n}}}}^{\gamma_{n}}}_{\gamma}
    {P}^{\gamma_{1} \gamma_{2} \cdots \gamma } \right), \qquad 
    \mbox{}\\
    \mbox{whence}  \quad 
     {P}^{\gamma_{1}\gamma_{2} \cdots \gamma_{n}}(T) =& 
\sum_{\delta_{i}} 
      { {M_{T}}^{\gamma_{1}} }_{\delta_{1}} 
     {{M_{T}}^{\gamma_{2}}}_{\delta_{2}} \cdots 
{{M_{T}}^{\gamma_{n}}}_{\delta_{n}}
      P^{\delta_{1} \delta \cdots \delta_{n}}, \quad \mbox{where}  
      \label{eq:PtensorFiniteEvolution} \\
      \quad \quad & { {M_{T{\mathbf k}}}^{\gamma_{i}} }_{\delta_{i}} 
=  
      {\left(e^{T  R_{ {\mathbf k}_{i}}}\right)^{\gamma_{i}} 
      }_{\delta_{i}}. \nonumber
\end{align}

It remains to transcribe these results into the path integral notation
(\ref{eq:PathIntRep}) and verify that the same time evolution is 
predicted in the `free' (Gaussian) case at least for time-independent 
rates. Clearly, for each degree of freedom ${\mathbf 
k}, \alpha$ there is a pair of classical paths ${\eta'(t)}_{{\mathbf 
k} \alpha},
{\eta(t)}^{{\mathbf k} \beta}$ or collectively simply $\eta'(t), 
\eta(t)$. From the fact that the rate operator 
is expressed by (\ref{eq:FreeRateOpMomSpace}) in normal form we take 
\begin{align}
    R_{t}(i\eta', \eta) = & \sum_{{\mathbf k}, \alpha, \beta}
    i{\eta'(t)}_{{\mathbf k} \alpha} {R_{{\mathbf 
k}}^{\alpha}}_{\beta}
    {\eta(t)}^{{\mathbf k} \beta}.\nonumber 
\end{align}
In these circumstances the time evolution kernel is particularly 
simple. 
Explicitly
\begin{align}
    \label{eq:PathIntRepGauss}
    {\mathcal M}_{T}(z', \zeta) = & \int {[}d\eta{]} {[}d \eta'{]} 
\exp{\left( \int_{0}^{T} dt ( -i\sum_{{\mathbf k}, \alpha} 
      \eta'{}^{{\mathbf k} \alpha}(t) 
\stackrel{\scriptscriptstyle{\bullet}}{\eta}_{{\mathbf k} 
      \alpha}(t)
     +i\sum_{{\mathbf k}, \alpha, \beta}
     {\eta'(t)}_{{\mathbf k} \alpha} {R_{\mathbf k}{}^{\alpha}}_{\beta}
     {\eta(t)}^{{\mathbf k} \beta}  ) + \sum_{{\mathbf k}, \alpha} 
z'{}_{{\mathbf k} \alpha} 
     \eta(T)^{{\mathbf k} \alpha} \right)}, 
         \end{align}    
subject to the appropriate boundary conditions. The integration over all paths
$\eta'(t)$, imposes a functional-$\delta$ constraint on $\eta(t)$, namely
\begin{align}
    \stackrel{\scriptscriptstyle{\bullet}}{\eta}^{{\mathbf k} 
\alpha}(t) =& \sum_{{\mathbf k},\beta}
     {R_{{\mathbf k}}^{\alpha}}_{\beta}{\eta(t)}^{{\mathbf k} \beta}, 
     \nonumber \\
 \mbox{whence }   \qquad    {\eta}^{{\mathbf k} \alpha}(T) =& \sum_{{\mathbf k},\beta}
     {M_{T}\mbox{}_{{\mathbf k}}^{\alpha}}_{\beta}{\eta(0)}^{{\mathbf 
k} \beta}
\end{align}
(if the transition rates are time-independent), where $M_{T {\mathbf 
k}}$ is given by (\ref{eq:PtensorEvolution}) above. Thus the $\eta(t)$ path integral contribution (up to a 
normalisation constant\footnote{
For this case the discrete version can be 
worked out explicitly as a (multidimensional) standard Gaussian 
integral, and the 
limit $N \rightarrow \infty$ considered. In the present heuristic 
discussion we simply assume that the steepest descent method yields 
the correct result.}) comes from the boundary term which 
gives using (\ref{eq:BoundaryConditions})
\begin{align}
   \label{eq:GaussianKernel}
   {\mathcal M}_{T}(z', \zeta) = & \texttt{C}e^{ \sum_{{\mathbf k} 
\alpha}
   {z'}_{{\mathbf k} \alpha} {M_{T}\mbox{}_{{\mathbf 
k}}^{\alpha}}_{\beta}
     {\zeta}^{{\mathbf k} \beta}         }
\end{align}   
for some integration constant \texttt{C}.

For a phylogenetic tensor state (as discussed above, with unit 
occupation number
in momentum modes ${\mathbf k}_{1},{\mathbf k}_{2}, \ldots, {\mathbf 
k}_{n}$) we set
\begin{align}
    P_{t}(z'_{1}, z'_{2}, \ldots, z'_{n}) =& 
    \sum_{\alpha_{1}, \alpha_{2}, \cdots \alpha_{n} =0}^{K-1} 
    P_{t}^{\alpha_{1}\alpha_{2} \cdots \alpha_{n}}z'_{{\mathbf k}_{1} 
\alpha_{1}} 
    z'_{{\mathbf k}_{2} \alpha_{2}} \ldots z'_{{\mathbf k}_{n} 
    \alpha_{n}}.
\end{align}
Denoting the state collectively as $P_{t}(z')$ we then have from 
(\ref{eq:OpAction})
\begin{align}
    \label{eq:IntegralFormTensorTimeEvol}
    P_{T}(z') = & \int \prod_{{\mathbf k}, \alpha}{[}d 
{\zeta'}_{{\mathbf  k} \alpha}{]} 
    {[}d {\zeta}^{{\mathbf k} \alpha}{]} {\mathcal 
    M}_{T}(z', \zeta) e^{-i \sum_{{\mathbf k}, \alpha} 
{\zeta'}_{{\mathbf  
    k} \alpha}{\zeta}^{{\mathbf k} \alpha}}P_{0}(i \zeta').
\end{align}
Substituting (\ref{eq:GaussianKernel}) it is evident that the $\zeta$ 
integrations impose a functional-$\delta$
constraint\footnote{Setting the arbitrary integration constant to 1.} 
$\delta(i\zeta' - z'\cdot M_{T})$, or 
\begin{align}
\label{eq:PtensorFiniteEvolution2}
   P_{T}(z') = & P_{0}(z'\cdot M_{T}), \quad \mbox{or} \nonumber \\
   P_{T}(z'_{1}, z'_{2}, \ldots, z'_{n}) =& 
    \sum_{\alpha_{1}, \alpha_{2}, \cdots \alpha_{n} =0}^{K-1} 
    P_{0}^{\alpha_{1}\alpha_{2} \cdots \alpha_{n}}
    {z'\cdot M_{T {\mathbf k}_{1}}}_{{\mathbf k}_{1} \alpha_{1}} 
     {z'\cdot M_{T {\mathbf k}_{2}}}_{{\mathbf k}_{2} \alpha_{2}} 
\ldots 
      {z'\cdot M_{T {\mathbf k}_{n}}}_{{\mathbf k}_{n} \alpha_{n}}, 
      \quad \mbox{or} \nonumber \\
    P_{T}(z'_{1}, z'_{2}, \ldots, z'_{n}) =& 
    \sum_{\alpha_{1}, \alpha_{2}, \cdots \alpha_{n} =0}^{K-1} 
    P_{T}^{\alpha_{1}\alpha_{2} \cdots \alpha_{n}}z'_{{\mathbf k}_{1} 
\alpha_{1}} 
    z'_{{\mathbf k}_{2} \alpha_{2}} \ldots z'_{{\mathbf k}_{n} 
    \alpha_{n}}, \quad \mbox{where finally} \nonumber \\
    {{P}_{T}}^{\gamma_{1}\gamma_{2} \cdots \gamma_{n}}=& 
\sum_{\delta_{i}} 
      { {M_{T}}^{\gamma_{1}} }_{\delta_{1}} 
     {{M_{T}}^{\gamma_{2}}}_{\delta_{2}} \cdots 
{{M_{T}}^{\gamma_{n}}}_{\delta_{n}}
      P_{0}^{\delta_{1} \delta \cdots \delta_{n}}  
\end{align}
as derived explicitly above.

It is instructive to 
re-write the classical `free' evolution kernel in 
Fourier transform space. Recalling the duality between `momentum' 
labels ${\mathbf k},{\mathbf l}  \in \pi {{\mathbb Z}_{2}}^{L}$ and 
`position' 
coordinates ${\mathbf x}, {\mathbf y} \in {{\mathbb Z}_{2}}^{L}$, 
define
\begin{align}
    \label{eq:PathIntegralRepXspaceForm}
    {\eta(t)}^{{\mathbf k} \alpha} = &  \sum_{{\mathbf x}}
    {\eta^{ \alpha}({\mathbf x},t)} e^{-i {\mathbf k} \cdot {\mathbf 
x}} , \qquad
    {\eta'(t)}_{{\mathbf k} \alpha} = {\frac{1}{2^{L}}}\sum_{{\mathbf 
y}}
    {\eta'_{\alpha}({\mathbf y},t)} e^{i{\mathbf k}\cdot {\mathbf 
y}}, \nonumber \\
    {R(t)_{{\mathbf k}}^{\alpha}}_{\beta} = & \sum_{{\mathbf z}}
    {R({\mathbf z},t)^{\alpha}}_{\beta}e^{i{\mathbf k}\cdot {\mathbf 
    z}}, \qquad \mbox{so that (\ref{eq:PathIntRepGauss}) becomes} 
\nonumber \\
    {\mathcal M}_{T}(z', \zeta) = & \int {[}d\eta{]} {[}d \eta'{]} 
\exp{\left( \int_{0}^{T} dt 
    ( -\!i \sum_{{\mathbf x}} {\eta'_{\alpha}({\mathbf x},t)} 
    {\stackrel{\scriptscriptstyle{\bullet}}{\eta}{\!\!}^{ 
\alpha}({\mathbf x},t)}
    \!+ \!i\!\sum_{{\mathbf x},{\mathbf y}}{\eta'_{\alpha}({\mathbf 
x},t)} {R({\mathbf x}\!-\!{\mathbf y},t)^{\alpha}}_{\beta}
    {\eta^{ \alpha}({\mathbf x},t)})+ \right.} \nonumber \\
    &\qquad \qquad  \qquad \qquad {\left. +\!
    \sum_{{\mathbf x}}{z'_{\alpha}({\mathbf x}}){\eta^{\alpha}({\mathbf x},T)}\right)}. 
\end{align}   
\section{Interaction terms and simple examples}
\label{sec:Interactions}
In this section we turn to the complete phylogenetic system 
incorporating
`interaction' terms.  In the previous section we constructed the 
`free'
part of the evolution kernel ${\mathcal M}_{T} = \int {[}d\eta{]} 
{[}d \eta'{]} 
\exp{S_{0}{[}\eta,\eta'{]}}$ for the phylogenetic `fields' 
${\eta'_{\alpha}({\mathbf
x},t)}$, ${\eta^{ \alpha}({\mathbf x},t)}$.  Incorporating
interactions, the kernel will acquire additional trilinear terms
$S_{1}$ in the exponent representing phylogenetic branching events, 
in such a way that
the manifest translation symmetry in `position' space is preserved.

In \S \ref{sec:Fock} above, it was pointed out that the `branching
operator' $\delta$ which was formally introduced in 
(\ref{eq:DeltaDefn}) of \S
\ref{sec:TensorMethods} can be represented by a trilinear $2
\leftarrow 1$ type operator in Fock space (compare
(\ref{eq:GammaTrilinear}) above).  In the case where there are up to
$L$ extant taxonomic units labelled by binary $L$-vectors (edge
`momenta') to allow for the development of a particular ancestral
binary tree, this vertex interaction must be given definite momentum
labels.  The labelling is of course always such that an edge ${\mathbf
k} \in {{\mathbb Z}_{2}}^{L}$ bifurcates to edges ${\mathbf l},
{\mathbf m}$ with ${\mathbf k}={\mathbf l}\!+\!{\mathbf m}$.  Moreover, if
the branching is pictured as a dynamical process, the interaction must
be time dependent.  The simplest possibility is that the system for $L$
taxa will evolve as a result of a fixed number $M$ of branchings at
times $t_{I}$, $I = 1,\ldots,M$ between times $t_{0}\equiv 0$ (from 
which time some assumed
ancestor(s) evolved) and $t_{M+1} \equiv T$ (the final time of 
measurement). 
A means of forcing these events is via $\delta(t-t_{I})$ interactions 
at times $t_{1}< \cdots < t_{I} < \cdots <t_{M}$ (with $0 \equiv 
t_0<t_1$ and $t_M < t_{M+1}\equiv T$).

With the above motivations we propose the following `interaction' 
term $S_{1}$
for the full evolution kernel ${\mathcal M}_{T} = \int {[}d\eta{]} 
{[}d \eta'{]} 
\exp{S{[}\eta,\eta'{]}}$ of the phylogenetic system, where $S = 
S_{0}+S_{1}$ with $S_{0}$ given by 
(\ref{eq:PathIntRepGauss},\ref{eq:PathIntegralRepXspaceForm}) and 
\begin{align}
\label{eq:InteractionTermAction}
    S_{1} = & - \int_{0}^{T} dt \frac 12 \sum_{I}\sum_{{\mathbf k}, {\mathbf 
l},{\mathbf m}}
    \delta(t-t_{I})\delta({\mathbf k}-{\mathbf l}-{\mathbf m})
    {\eta'(t)}_{{\mathbf l} \alpha} {\eta'(t)}_{{\mathbf m} \beta}
    {\Gamma^{\alpha \beta}}_{\gamma} {\eta(t)}^{{\mathbf k} \gamma} 
    \nonumber \\
    =& - \int_{0}^{T} dt \frac 12 \sum_{I}\sum_{{\mathbf x}} \delta(t-t_{I})
    {\eta'({\mathbf x},t)}_{\alpha} {\eta'({\mathbf x},t)}_{ \beta}
    {\Gamma^{\alpha \beta}}_{\gamma} {\eta({\mathbf x}, t)}^{\gamma}. 
\end{align}
As expected, the binary edge labelling is reflected in the manifest 
translation symmetry of this expression.
With the complete model $S_{0}+S_{1}$, the path integral formalism can
now be used to construct (in a perturbation expansion,
see below) the evolution kernel for the
full system, and hence transition probabilities for evolution, from
any initial state to any final state.  In the case of phylogenetic
inference, one is of course interested in evolution from an initial
root edge (at time $t=0$) to (at time $t=T$) an observed joint
probability density for character types of $L$ taxonomic units.  

The model (\ref{eq:PathIntRepGauss}), 
(\ref{eq:PathIntegralRepXspaceForm}),
(\ref{eq:InteractionTermAction}) is \textit{generic} in the sense that
an \textit{arbitrary} (but fixed) number of branching events $M$, and
\textit{any} compatible branching processes for binary edges, is
encoded.  If probabilities 
For connected binary trees, with a single root and $L$ leaves one should of course admit only 
$M=L-1$ $\delta$-function forcing terms, and adopt
standard momentum labelling, for example for $L \leftarrow 1$ the root
edge may be chosen as the binary $L$-vector
$(1,1,\ldots,1)$, and the edges the binary
$L$-vectors $(0,0,\ldots,1)$, $(0,\ldots,1,0)$, $(0,\ldots,1,0,0)$,
$\cdots$ (denoted below by decimal equivalents $\vec{1}$, $\vec{2}$, 
$\vec{4}$, $\cdots$). For formal
analysis with a specific tree, it may in fact be combinatorially more
powerful to consider \textit{all} such admissible $L \leftarrow 1$
momentum routing schemes.

For completeness, we derive in the appendix, \S A, a formal
perturbative expansion\cite{Peliti1985},  and give explicit Feynman rules for the present
model (see table \ref{tab:FeynRules}).  The evolution
kernel for $S_{0}+S_{1}$ is re-written by expanding 
$\exp({S_{1}})$ in a power series, so that the essential ingredients 
are specific path integrals of monomials in the phylogenetic path
variables, weighted by the `free' part. In turn, these moments 
can be reduced to functional derivatives of  an extended  `free'  kernel, 
with respect to ancillary `external' path variables coupled by additional linear terms to the 
path variables which are being integrated over.
The extended kernel is again quadratic and can be evaluated as a Gaussian
in terms of the formal inverse bilinear form or propagator with appropriate boundary conditions 
(see appendix \S A).
Moreover, the $\delta$-function forcing terms require the derivatives with 
respect to the external path variables to be
evaluated at the interaction times $t_{I}$.  The probabilities (pattern frequencies corresponding to
\textit{all} binary $L$ leaf trees with evolution on edges determined
by the specific edge rates $R_{\mathbf k}(t)$) so constructed are \textit{identical}
to the usual likelihood calculation via extended leaf colourations for example.
In the earlier second-quantized version,
(see \cite{BashfordJarvis2001}), the model was constructed using 
the canonical (creation and annihilation operator) formalism, and the
interaction term treated in time dependent perturbation theory.
We emphasize that, although well-known, the result in our formalism follows
\textit{automatically} from the time evolution kernel for the model
(effectively, an appropriate Markov rate operator lifted to the whole
Fock space), so that in this sense we have produced a truly dynamical
model for phylogenetic branching processes.

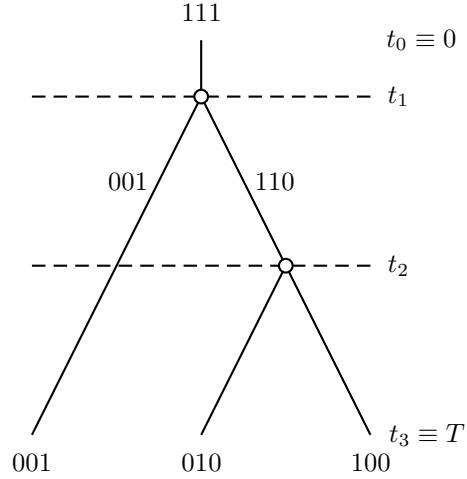
\begin{figure}[tbp]
		\centering
\raisebox{.5cm}{\begin{pspicture}(6,6)
\psset{framearc=0.0,framesep=0.0truecm,nodesep=8pt,unit=0.75truecm,
      arrowsize=4pt 2,arrowinset=0.4}
\def\pstlw{0.8pt}      
\psline{-}(0,0)(3,6)
\psline{-}(6,0)(3,6)
\psline{-}(3,6)(3,7)
\psline{-}(3,0)(4.5,3)
\psline[linestyle=dashed]{-}(0,3)(6,3)
\psline[linestyle=dashed]{-}(0,6)(6,6)
\pscircle[linewidth=0.8pt,fillstyle=solid,fillcolor=white](3,6){.14} 
\pscircle[linewidth=0.8pt,fillstyle=solid,fillcolor=white](4.5,3){.14}  
\rput(6.9,7){$t_0 \equiv 0$}
\rput(6.5,6){$t_1$}
\rput(6.5,3){$t_2$}
\rput(7,0){$t_3 \equiv T$}
\rput(3,7.5){$111$}
\rput(1.7,4.5){$001$}
\rput(4.3,4.5){$110$}
\rput(0,-.5){$001$}
\rput(3,-.5){$010$}
\rput(6,-.5){$100$}
\end{pspicture} }
	\caption{Binary labelling scheme for a tree on 3 leaves
	${\mathcal T}= (\vec{1}(\vec{2}\vec{4}))$ with branching events at 
intermediate times $t_{1}$,
	$t_{2}$.  Nonzero rate constants for the model are chosen for
	the root and leaf momenta $\vec{7} = (111)$, $\vec{1} =
	(001)$, $\vec{2}=(010)$ and $\vec{4}=(100)$ respectively,
	together with a \textit{single} additional momentum
	$\vec{6}=(110)$.}
	\label{fig:TreeFig}
\end{figure}

We illustrate our results by reiterating some concrete examples
from \cite{BashfordJarvis2001} together with some further remarks.
Consider the case $L=3$, $M=2$.  Nonzero rate constants are chosen for
the root and leaf momenta $\vec{7} = (111)$,
$\vec{1} = (001)$, $\vec{2}=(010)$ and
$\vec{4}=(100)$ respectively, together with a
\textit{single} additional momentum $\vec{6}=(110)$
associated with the tree ${\mathcal T}= (\vec{1}(\vec{2}\vec{4}))$ of 
figure
\ref{fig:TreeFig}.  Clearly the contribution to the $3 \leftarrow 1$
scattering probability (or likelihood) associated with this tree is, 
as required, the
term arising (in the operator formalism \cite{BashfordJarvis2001}) 
from inserting intermediate
states in the above with the correct intermediate edge momenta, or (in
the perturbation expansion of the path integral method) from the
correct linking of propagators and vertices at this order (see Feynman rules in appendix, \S A, and table \ref{tab:FeynRules}).  
\textit{Either}
approach gives finally
\begin{align}
\label{eq:ScatteringProbCalc}
{P_{\mathcal T}}^{\alpha_{\vec{1}} \alpha_{\vec{2}}
	\alpha_{\vec{4}}} = & {\langle 
\alpha_{\vec{1}} {\vec{1}}, \alpha_{\vec{2}} {\vec{2}}, 
	\alpha_{\vec{4}} {\vec{4}}\;|P_{\mathcal T}(T) \rangle} = {\langle 
\alpha_{\vec{1}} {\vec{1}}\; \alpha_{\vec{2}} {\vec{2}}\; 
	\alpha_{\vec{4}} {\vec{4}}\;|
	M_{\mathcal T}(T,0)|\;p_{{\vec{7}}}(0)\rangle } \nonumber \\
	=& \sum { M_{\vec{2}}{}^{\alpha_{\vec{2}}} }_{\beta_{\vec{2}}	}
{M_{\vec{4}}{}^{\alpha_{\vec{4}}} }_{\beta_{\vec{4}}}
{\Gamma_2{}^{\beta_{\vec{2}} \beta_{\vec{4}} }}_{\gamma_{\vec{6}}}\cdot 
 { M_{\vec{6}}{}^{\gamma_{\vec{6}}} }_{\beta_{\vec{6}}}
{ M_{\vec{1}}{}^{\alpha_{\vec{1}}} }_{\beta_{\vec{1}}} 
{\Gamma_{1}{}^{\beta_{\vec{1}} \beta_{\vec{6}}}}_{\beta_{\vec{7}}} 
\cdot 
  { M_{\vec{7}}{}^{\beta_{\vec{7}}} }_{\alpha_{\vec{7}}}  
p^{\alpha_{\vec{7}}}.
\end{align}
Here $ |\;p_{{\vec{7}}}(0)\rangle  =  
\sum p^{\alpha_{\vec{7}}}(0) |\alpha_{\vec{7}} {\vec{7}} \rangle$ is 
the 
state representing the initial root edge probability density, and the 
$M_{\mathbf k}$ are the Markov transition matrices for the 
appropriate 
edges, namely $M_{\mathbf k} = e^{\Delta_{{\mathbf k}}R_{{\mathbf 
k}}}$ with $\Delta_{{\mathbf k}}$ the time interval for evolution on 
edge ${\mathbf k}$, $\Delta_{{\mathbf k}} = t_{I'}-t_{I}$ where the 
branching 
times at the source and target of edge ${{\mathbf k}}$ are $t_{I}$
and $t_{I'}$.

\begin{figure}[tbp]
		\centering
\raisebox{.5cm}{\begin{pspicture}(4.5,8)
\psset{framearc=0.0,framesep=0.0truecm,nodesep=8pt,unit=0.75truecm,
      arrowsize=4pt 2,arrowinset=0.4}
\def\pstlw{0.8pt}      
\psline{-}(0,3)(1.5,6)
\psline{-}(1.5,3)(1.5,6)
\psline{-}(1.5,6)(4.5,0)
\psline{-}(1.5,6)(1.5,7.5)
\psline[linestyle=dashed]{-}(0,5.8)(3,5.8)
\psline[linestyle=dashed]{-}(0,6.2)(3,6.2)
\pscircle[linewidth=0.8pt,fillstyle=solid,fillcolor=white](1.5,6){.14} 
\rput(3.2,6.5){$t_1$}
\rput(3.2,5.5){$t_2$}
\rput(1.5,8){$111$}
\rput(0,2.5){$001$}
\rput(1.5,2.5){$010$}
\rput(4.5,-.5){$100$}
\end{pspicture} }\hspace*{.5cm}
\raisebox{.5cm}{\begin{pspicture}(3,8)
\psset{framearc=0.0,framesep=0.0truecm,nodesep=8pt,unit=0.75truecm,
      arrowsize=4pt 2,arrowinset=0.4}
\def\pstlw{0.8pt}      
\psline{-}(0,3)(1.5,6)
\psline{-}(1.5,0)(1.5,6)
\psline{-}(1.5,6)(3,3)
\psline{-}(1.5,6)(1.5,7.5)
\psline[linestyle=dashed]{-}(0,5.8)(3,5.8)
\psline[linestyle=dashed]{-}(0,6.2)(3,6.2)
\pscircle[linewidth=0.8pt,fillstyle=solid,fillcolor=white](1.5,6){.14} 
\rput(3.2,6.5){$t_1$}
\rput(3.2,5.5){$t_2$}
\rput(1.5,8){$111$}
\rput(0,2.5){$001$}
\rput(1.5,-.5){$010$}
\rput(3,2.5){$100$}
\end{pspicture} }
	\caption{Additional effective \textit{non-binary} trees
	${\mathcal T}_{\vec{3}}$ and ${\mathcal T}_{ \vec{5}}$
	contributing to the probability in the phylogenetic branching
	model for the three leaf case.  Non-propagating momenta
	$\vec{3} = (011)$ and $\vec{5} = (101)$ produced by the
	branching interaction term at $t_{1}$ cause effective
	trivalent vertices with different evolution times $T-t_{1}$,
	$T-t_{2}$ on long and short leaf edges.}
	\label{fig:TriTreeFig}
\end{figure}
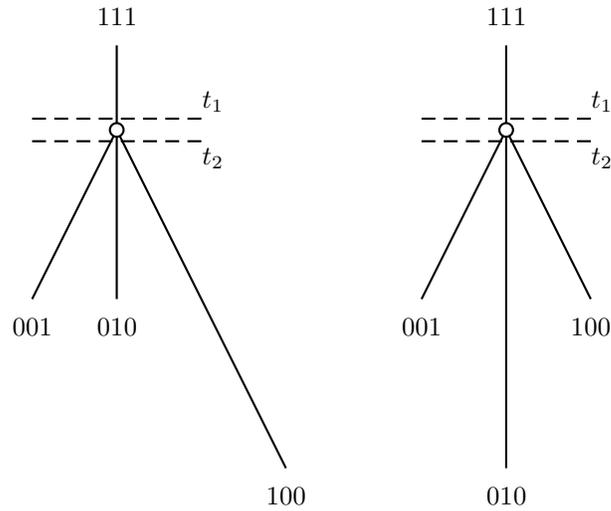

As indicated by the ${\mathcal T}$ subscript in
(\ref{eq:ScatteringProbCalc}), the total expression for $P(T)$
includes terms additional to the contribution from the selected tree. 
In fact without additional subtraction terms (see
\cite{Doi1976a,Doi1976b,Peliti1985,BashfordJarvis2001},) the model as formulated
is not probability conserving.  However, in phylogenetic inference
(for example maximum likelihood analyses) it is appropriate to
generate contributions from \textit{all} candidate trees for unknown
rates.  In the present case the additional terms arise of course from other admissible trees.
In fact, even if only the rate constants for edges specific to a selected tree are nonzero,
there are still contributions (in the
operator approach) from intermediate states with non-propagating
momenta, and these also arise in the combinatorics of the path integral
perturbation expansion(see below).  Thus in
addition to (\ref{eq:ScatteringProbCalc}) there are the trees with
effective \textit{trivalent} nodes,
\begin{align}
    \label{eq:TrivalentI}
    P_{{\mathcal T}_{\vec{3}}} = & \sum { 
M_{\vec{1}}{}^{\alpha_{\vec{1}}} }_{\beta_{\vec{1}}	}
    { M_{\vec{2}}{}^{\alpha_{\vec{2}}} }_{\beta_{\vec{2}}	}  
     { M'_{\vec{4}}{}^{\alpha_{\vec{4}}} }_{\beta_{\vec{4}} }
     {\Gamma^{\beta_{\vec{1}}\beta_{\vec{2}} 
\beta_{\vec{4}}}}_{\gamma_{\vec{7}}}\cdot 
     { M_{\vec{7}}{}^{\gamma_{\vec{7}}} }_{\beta_{\vec{7}}}  
     p^{\beta_{\vec{7}}} \\ 
    P_{{\mathcal T}_{\vec{3}}} = & \sum { 
M_{\vec{1}}{}^{\alpha_{\vec{1}}} }_{\beta_{\vec{1}}	}
    { M'_{\vec{2}}{}^{\alpha_{\vec{2}}} }_{\beta_{\vec{2}}	}  
     { M_{\vec{4}}{}^{\alpha_{\vec{4}}} }_{\beta_{\vec{4}} }
     {\Gamma^{\beta_{\vec{1}}\beta_{\vec{2}} 
\beta_{\vec{4}}}}_{\gamma_{\vec{7}}}\cdot 
     { M_{\vec{7}}{}^{\gamma_{\vec{7}}} }_{\beta_{\vec{7}}}  
     p^{\beta_{\vec{7}}}      \label{eq:TrivalentII}
\end{align}
shown in figure \ref{fig:TriTreeFig}.  The \textit{tri}valency comes
by deleting edges for rates with $R_{\mathbf k}=0$ and re-joining the
target and source nodes such that there is an effective 3 point vertex
corresponding to a branching operator or interaction vertex structure coefficient 
with components (compare
(\ref{eq:GammaMatrixForm})) ${\Gamma_{\alpha}}^{\beta \gamma \delta} =
{\delta_{\alpha}}^{\beta}{\delta_{\alpha}}^{\gamma}{\delta_{\alpha}}^{\delta}$
.  (Such an effective interaction term might also be viewed as the
result of directly integrating out the $\eta^{\mathbf k},
\eta'_{\mathbf k}$ variables corresponding to $R_{\mathbf k}=0$).  For
the tree in question, the non-propagating momenta are $\vec{3} =
(011)$ and $\vec{5} = (101)$ corresponding to the trees ${\mathcal
T}_{\vec{3}}$ and ${\mathcal T}_{ \vec{5}}$ respectively.  The terms
differ from one another because the edge evolution times $T-t_{1}$ and
$T-t_{2}$ are distributed differently over the Markov matrices
$M_{\vec{1}}$, $M_{\vec{2}}$ and $M_{\vec{4}}$ as indicated by the $^\prime$ in (\ref{eq:TrivalentI}),
(\ref{eq:TrivalentII})  and the 
differing edge lengths in figure \ref{fig:TriTreeFig}. Of course, it is always possible to regard these terms as 
vestigial contributions from standard \emph{binary} trees with very short edges. In fact, since
the usual counting relation between edges and leaves for binary trees
obviously does not hold for the trivalent trees, the formal introduction of a scaling
parameter would serve to distinguish these and similar noncanonical
tree diagrams.
\section{Conclusions and discussion}
\label{sec:Concl}
In this paper and the previous work \cite{BashfordJarvis2001} we have
proposed a transcription of phylogenetic branching processes into the
language of a stochastic dynamical system evolving according to an
appropriate Markov rate operator on a suitably extended `state' 
space. 
The analogy with statistical and particle physics is that the
`particles' in the phylogenetic context are the individual taxonomic
units, and it is these which evolve in type and number (as in Markov
models of reaction diffusion or birth-death processes, or in 
relativistic particle scattering) in the
course of evolution.  In \cite{BashfordJarvis2001} a conventional
operator approach was taken, whereas in the present work the path
integral formulation introduces to phylogenetics the familiar 
physical notions of
`paths' and `fields' (over a discrete
 lattice).  Our treatment including
`interactions' representing branching events, including explicit Feynman rules,
(table \ref{tab:FeynRules} in appendix, \S A) establishes the equivalence of the path integral
formulation to the operator version \cite{BashfordJarvis2001} via standard perturbation
theory as the appropriate tool for completing the transcription.

The path integral language allows a range of techniques known in the
context of the analysis of physical systems \cite{Peliti1985,Doi1976a,Doi1976b}
to be deployed for phylogenetics.  One immediate point is the
relationship between the formulation of transition probabilities in
`momentum' space versus the dual `position' space - standard in
condensed matter systems, and also known in phylogenetics in the
literature on transform techniques for phylogenetic inference 
involving the
discrete Fourier Hadamard transformation, in principle to derive an
edge rate spectrum for a phylogenetic tree directly from an observed
data set of pattern probabilities
\cite{SzekeleyErdosSteelPenny1993,HendyPennySteel1994,SteelHendyPenny1998}. 
In the present framework, momentum conservation is a reflection of 
translation invariance on the underlying discrete lattice.

General considerations for the path integral formulation include the
further application of symmetry principles in various ways. For example,
\textit{continuous} Lie symmetry group
actions on the path variables (phylogenetic fields), for
example $\bar{\eta} \rightarrow \bar{\eta}\cdot \overline{U}$, ${\eta}
\rightarrow U \cdot {\eta}$ can be analysed for their
effect on the dependence of the time evolution kernel on the various
rate and time parameters of the model.  This has been pursued in
\cite{BashfordJarvisSumnerSteel2004} (in the explicit tensor
description) for the well known Kimura 3ST model for 4 characters
\cite{Kimura1981} where it was noted that the rate and branching
operators intertwine the action of a $U(1)\times U(1)\times U(1)$ ( or
${\mathbb C}^{\times}\times {\mathbb C}^{\times}\times {\mathbb
C}^{\times}$) group so that the resultant group reduction from
representations of $SU(4)$ (or $SL(4)$) is intimately related to the
properties of this model (it is well known that the Kimura 3ST model
and the related 2P model do belong to the class of \textit{discrete}
colour group models \cite{SteelHendyPenny1998,SempleSteel2003}).  More
generally, the Lie symmetry approach allows rate models to be
considered in principle in terms of a hierarchy of symmetry-breaking
terms.  For example, in molecular phylogenetics at the protein
mutation level, suitable symmetry groups would be those advocated
recently in relation to the possible group-theoretical origins of the
genetic code itself (see for example
\cite{BashfordTsohantjisJarvis1998,BashfordJarvis2000}).

A deeper aspect of the branching model in the path integral
formulation is the role of time reparametrisations, $t \rightarrow
\tau(t)$, in connection with notions of the `molecular clock'.  Given
that
\begin{align}
    dt = d\tau \frac{dt}{d\tau}, \quad 
    & \quad \delta(t - t_{I}) dt = 
    \frac{\delta(\tau - \tau_{I})}{|{dt}/{d\tau}|}\cdot 
\frac{dt}{d\tau} d\tau, \nonumber
\end{align}
then clearly the evolution kernel has
the following covariance property,
\begin{align}
    \label{eq:TimeReparametrisationCovariance}
    {\mathcal M}_{T}(t_{I}, R_{\mathbf k}(t)) =&\, {\mathcal 
    M}_{T}(\tau_{I}, R'_{\mathbf k}(\tau)), \quad \mbox{where} \quad  
    R'_{\mathbf k}(\tau) = R_{\mathbf k}(t)\cdot \frac{dt}{d\tau}
    \quad \mbox{and} \quad \tau_{I} = \tau(t_{I})
\end{align}
(it is assumed that $dt/d\tau >0$, in particular $\tau(t)$ is not orientation reversing).  This  is precisely the reason
that, in standard probability approaches (see for example
\cite{SteelHendyPenny1998}), `dynamical' considerations involving
explicit time dependence can be absent -- standard calculations
require only the combinatorics of the tree (which is encoded in the
present models via the branching times $t_{I}$ and the choice of
momenta for which rates are nonzero).  However, as has been mentioned
already, there is good reason to formulate the branching process
temporally as presented here.  In order for generalisations of the 
$\delta$-function forcing interaction terms to
preserve the time reparametrisation covariance
(\ref{eq:TimeReparametrisationCovariance}), the introduction of an
auxiliary phylogenetic `gauge'  field would be mandatory (as
in some proper time formulations of relativistic field equations).

As an illustration of this dynamical perspective, suppose now that
for some edge momentum ${\mathbf k}^{*}$ the edge rate can be written
as proportional to some standard rate matrix,
\begin{align}
    \label{eq:MultiplierRateDefn}
    R_{{\mathbf k}^{*}}(t) =&  \lambda^{*}(t) \cdot R^{*}.
\end{align}
Then it is possible to define a phylogenetic `proper time' $\tau^{*}$
(implicitly) as a function $t$, by solving the first order equation
\begin{align} 
    \frac{dt}{d\tau^{*}} = & \frac{1}{\lambda^{*}(t)} \nonumber
\end{align}
together with some suitable initial condition, for example
${\tau^{*}}_{I}={\tau^{*}}(t_{I}) \equiv t_{I}$ where $t_{I}$ is the
branching time at the source node of edge ${\mathbf k}^{*}$.  Then, 
with respect
to this proper time, the edge rate $R_{{\mathbf k}^{*}}(\tau^{*})$ is
by definition \textit{constant}, and equal to ${R^{*}}$.  By
extension, if there exists a distinguished \textit{tree path} ${\mathcal
P}^{*}$ from the root to some leaf node, along which \textit{all} edge
rate matrices possess the above multiplier property
(\ref{eq:MultiplierRateDefn}), a \textit{global} phylogenetic proper
time $\tau^{*}$ exists for that tree path, with the rate matrices piecewise
constant (constant along each edge).  Finally, such a tree path
phylogenetic proper time may always be adjusted to \textit{coincide}
with geological or archaeological time determinations at certain
points by piecewise linear affine transformation(s) of the form
$\tau^{*} \rightarrow a\tau^{*} + b$ (which may be edge dependent 
along the distinguished path) without compromising the
above arguments.  An extreme example of this situation is of course
the case of a stationary Markov process, wherein \textit{each} rate
matrix is (proportional to) a given fixed matrix $R$, 
and the (weighted) sum of elapsed times along each tree path from the root to a leaf node, 
is constant  -- in this case a molecular clock exists in the strongest sense.  As
usual however, it is still impossible to disentangle evolution
occurring on some edges with standard strength for time $\Delta
t$, from evolution occurring over time $\lambda \Delta t$ with
scaled rates $(\lambda^{-1})R$.  In general, 
conclusions drawn from studies of `time dependent' rate matrices
should always be treated with caution because of reparametrisation
covariance.  Related considerations for general Kolmogorov 
equations, related to non-stationary finite Markov processes, 
but without explicit recognition of the role of the 
group of time reparametrisations (diffeomorphisms), have 
been given in \cite{Goodman1970};  for a discussion of general time-dependent 
Markov processes see \cite{RindosWooletViniotisTrivedi1995}. The `intrinsic time' of 
\cite{Goodman1970} is nothing but the above phylogenetic `proper time' $\tau$. This, in turn 
-- interpreted as a gauge fixing choice -- is essentially the Teichm\"{u}ller parameter for the 
configuration space of an implicit `einbein' path variable which carries 
gauge transformations associated with the 
group of time reparametrisations on the interval $[0,T]$ 
(see for example \cite{Govaerts1991}).

Within the present reformulation it is also possible to examine
generalisations which may not be apparent in other contexts.  An example
would be analytical or at least systematic possibilities for the examination 
of the behaviour random trees in the limit of very large numbers of leaves, 
or of random branching events, for the purpose of comparative evolution studies.  
A further extension would be to include population processes such as 
mutation-selection effects into the models.

A final potentially important analytical tool is the fact that (as
mentioned above) the closed form expression for the scattering
probabilities represented by the evolution kernel generates contributions 
from \textit{all} candidate trees for a
given number of leaves.  It is clear from our
presentation that the characteristics of a specific tree can be
encoded via the choice of nonzero rate constants for particular edge
momenta, and that there may be several equivalent such assignments
amongst the $2^{L}$ admissible binary $L$-vectors.  The exploitation
of the interrelations of these assignments might give insights into
the derivation of `invariants' (in this case for the combinatorics of
trees, rather than for differential topology, as in the case of
topological quantum field theory) which could provide useful
constraints in phylogenetic inference and maximum likelihood `optimal'
tree searches. Indeed, in maximum likelihood approaches themselves, it 
may be useful to have a formal representation of all contributing terms, 
without the need for explicit tree enumerations.

\paragraph{Acknowledgements:} PDJ and JGS thank the Department of
Physics and Astronomy, and also the Biomathematics Research Centre,
University of Canterbury, Christchurch, New Zealand, for hosting a
visit in 2003 during which this work was presented.  PDJ also 
acknowledges the hospitality of the Biosciences Research Centre, 
Massey University, during an earlier visit. This research was
supported by the Australian Research Council grant DP0344996.
We thank the anonymous referees for suggestions for strengthening 
an earlier version.

\vfill

{\small
 

}
\pagebreak
\begin{appendix}

\section{\label{sec:FeynRules}Feynman rules for phylogenetic branching models}
\setcounter{equation}{0}
\renewcommand{\theequation}{{A}-\arabic{equation}}
In this appendix we develop systematic expansion methods in the form of Feynman rules, for the calculation of the time evolution of state probabilities in the model given by (\ref{eq:PathIntRepGauss}) and (\ref{eq:PathIntegralRepXspaceForm}). This establishes that the model 
is formally equivalent to the standard prescription for calculating likelihood functions for phylogenetic trees, and provides the justification for the more qualitative discussion of the free and interacting cases given in \S\S \ref{sec:FreePathInt} and 
\ref{sec:Interactions} above.

Firstly note that the path integral representation of the (free) time evolution kernel ${\mathcal M}^\circ_T(z, \zeta)$,  (\ref{eq:PathIntRep}), (\ref{eq:PathIntRepGauss}) and (\ref{eq:PathIntegralRepXspaceForm}) can be written in various equivalent symmetrised forms emphasising the role of the boundary conditions, namely (using the generic form (\ref{eq:PathIntRep}) to suppress affixes)\footnote{Using either of the second two forms in the discussion following (\ref{eq:PathIntRepGauss}) leads to  equivalent solutions, for example 
\begin{align}
i\eta'_{{\mathbf k}\alpha}(T)= & \, i\eta'_{{\mathbf k}\beta}(0)  
{(M_{{\mathbf k}T}{}^{-1}){}^{\beta}}_{\alpha}, \quad 
\mbox{and then} \quad  i \eta'(0) \zeta =  i\eta'_{{\mathbf k}\beta}(0)  
{(M_{{\mathbf k}T})^{\beta}}_{\alpha} \zeta^\alpha  \nonumber 
\end{align}
as before.}, 
\begin{align}
\label{eq:SymmetrisedForms}
 \int_{0}^{T} dt \left( -i 
\eta'(t)\stackrel{\scriptscriptstyle{\bullet}}{\eta}(t)\right. +  & \left.  R_{t}(i 
\eta', \eta) \right)  +{z \eta(T)}  =    \int_{0}^{T} dt \left( +i 
\stackrel{\scriptscriptstyle{\bullet}}{\eta}{\!\!\!}'(t)\eta(t) + R_{t}(i 
\eta', \eta) \right) + i \eta'(0) \zeta \nonumber \\
 & \, = \int_{0}^{T} dt \left( -i \frac 12(\eta'(t)\stackrel{\scriptscriptstyle{\bullet}}{\eta}(t)
 - \stackrel{\scriptscriptstyle{\bullet}}{\eta}{\!\!\!}'(t)\eta(t)) + R_{t}(i \eta', \eta) \right) 
    +\frac 12 {z \eta(T)} + \frac 12 i \eta'(0)\zeta.
\end{align}
Now consider the complete time evolution kernel extended by some ancillary path variables $i \xi'(t), \xi(t)$,
\begin{align}
\label{eq:SymbolicNotation}
\widetilde{\mathcal M}_{T} :=  & \, \int d{[}\eta'{]} d{[}\eta{]} \exp \left(i \eta' \cdot K \cdot \eta + i\xi' \cdot \eta +
                                        i\eta'\cdot \xi \right) \exp(+{z \eta(T)}) \exp S_1{[}i \eta', \eta{]},
\end{align}
such that $\widetilde{M}_{T} \stackrel {\xi'=0= \xi}{\longrightarrow} {M}_{T}$. The  notation `$\cdot$' in the exponential
represents a definite integral of the occurring path variables with respect to time from $t=0$ to $t=T$, respecting of course the boundary conditions derived earlier, (\ref{eq:BoundaryConditions})\footnote{
Bearing in mind that the additional boundary contributions are for specific times, and are thus products, \emph{not} integrals.}. The notation  ` $i\eta' \cdot K \cdot \eta$'  refers to the quadratic part of the integrand, in this case in the first of the forms (\ref{eq:SymmetrisedForms}).
Finally an additional (for the moment generic)  `interaction'  term 
is included, with $S_1$ being the integral of the normal kernel. 

The aim is to consider the convolution of 
$\widetilde{M}_{T}$ with the initial state probability generating function, in such a way that 
the the expansion of the exponential of the interaction in a power series, together with the final state matrix element, and  the folding with respect to the initial state probability tensor,  are all reduced to formal derivatives with respect to the ancillary variables, acting on the expression for the `free' kernel, 
\begin{align}
\widetilde{\mathcal M}^\circ_{T}(z,\zeta) :=  & \, \int d{[}\eta'{]} d{[}\eta{]} \exp \left(i \eta' \cdot K \cdot \eta + i\xi' \cdot \eta + i\eta' \cdot \xi \right) \exp(+{z \eta(T)}).
\end{align}

To this end consider the complete generating function for the final probability state vector (compare
(\ref{eq:OpAction})),
\begin{align}
\label{eq:FinalStateVectorFull}
P_T(z) = & \, \left.  \int   d\zeta d\zeta' \widetilde{\mathcal M}_{T}(z,\zeta)e^{-i \zeta'\zeta}P_0(\zeta')\right|_{\xi \equiv \xi' \equiv 0}.
\end{align}
The additional $S_1{[}\eta, i\eta'{]}$ interaction term in the exponential can be regarded, after a power series expansion, as a series of moments evaluated on the free kernel, so that
\begin{align}
\label{eq:FinalStateVectorExpansion}
P_T(z) = & \, \left. e^{\displaystyle {S_1{[}\frac{\partial}{\partial {i\xi'}}, \frac{\partial}{\partial \xi } }{]}}\cdot \int   d\zeta d\zeta' \widetilde{\mathcal M}^\circ_{T}(z,\zeta)e^{-i \zeta'\zeta}P_0(\zeta')\right|_{\xi \equiv \xi' \equiv 0}.
\end{align}
Also if we are interested in a final state consisting of $L$ taxonomic units, the relevant probability component is by definition the generating function derivative with respect to the appropriate $z$ variables; for example
\begin{align}
P_T^{\alpha_1 {\mathbf k}_1 \dots \alpha_L {\mathbf k}_L} = & \,
          \left.  \frac{\partial}{\partial z_{\alpha_1 {\mathbf k}_1}} \cdots 
           \frac{\partial}{\partial z_{\alpha_L {\mathbf k}_L}}  P_T(z) \right|_{z \equiv 0}
\end{align}
From the dependence of the kernel on $z$ (the first form in (\ref{eq:SymmetrisedForms})), however, the derivatives merely bring down factors of $\eta(T)$ with appropriate labels, which in turn are equivalent to the corresponding differentiations with respect to $i\xi'$:
\begin{align}
P_T^{\alpha_1 {\mathbf k}_1 \dots \alpha_L {\mathbf k}_L} = & \,
          \left.  \exp {S_1 \Big[ \frac{\partial}{\partial {i\xi'}}, \frac{\partial}{\partial \xi }} \Big]  \cdot \frac{\partial}{\partial i\xi'_{\alpha_1 {\mathbf k}_1}} \cdots 
           \frac{\partial}{\partial i\xi'_{\alpha_L {\mathbf k}_L}} \cdot 
           \int   d\zeta d\zeta' \widetilde{\mathcal M}_{T}(z,\zeta)e^{-i \zeta'\zeta}P_0(i\zeta')
             \right|_{z \equiv \xi \equiv \xi'=0}
\end{align}
Finally, from the \emph{second} form of the kernel in (\ref{eq:SymmetrisedForms}), the path integral over $\zeta$ will enforce a $\delta$-function constraint identifying $i\zeta' $ with $i\eta'(0)$, or partial differentiation with respect to the appropriate components of $\xi$:
\begin{align}
\label{eq:FinalProbabilityComponent}
P_T^{\alpha_1 {\mathbf k}_1 \dots \alpha_L {\mathbf k}_L} = & \,
        \left.  \exp {S_1\Big[  \frac{\partial}{\partial {i\xi'}}, \frac{\partial}{\partial \xi  }} \Big]   \cdot \frac{\partial}{\partial   i\xi'_{\alpha_1 {\mathbf k}_1}(T)} \cdot 
   \frac{\partial}{\partial i\xi'_{\alpha_L {\mathbf k}_L}(T)} \cdot 
        P_0\left( \frac{\partial}{\partial \xi(0)} \right) \cdot \widetilde{\mathcal M}^\circ_{T}(z,\zeta )
             \right|_{z \equiv \xi \equiv \xi'=0}
\end{align}

Turning to the evaluation of $\widetilde{\mathcal M}^\circ_{T}(z,\zeta )$ itself, note that the quadratic part of the integrand in (\ref{eq:SymbolicNotation}) can be written
\begin{align}
\label{eq:QuadraticIntegrand}
 \int \hspace*{-.2cm} \int_0^T dt dt'  i\eta' (t')\cdot K(t,t') \eta(t)  = & \, \int \hspace*{-.2cm} \int_0^T dt dt' i\eta'(t) (-\partial_t \delta(t-t') + R \delta(t-t')) \eta(t'). 
 \end{align}
The formal completion of the square 
\begin{align}
i \eta' \cdot K \cdot \eta + i\xi' \cdot \eta +
                                        i\eta'\cdot \xi = & \, i(\eta' + \xi' K^{-1})\cdot K\cdot  (\eta + K^{-1} \xi) - 
                                        i \xi' \cdot K^{-1}\cdot \xi
\end{align}
suggests integrating out the resulting Gaussian after the change of variables $i\eta' \rightarrow  i(\eta' + \xi' K^{-1})$,
$\eta \rightarrow (\eta + K^{-1}\xi)$, (which has unit Jacobian), leaving  the expression
\begin{align}
\label{eq:FreeKernelExplicit}
\widetilde{\mathcal M}^\circ_{T}(z,\zeta) = \exp( - i\xi' \cdot K^{-1}\cdot  \xi)
\end{align}
up to normalisation factors ( including $\det{K}^{-1}$) and boundary contributions. However, the explicit dependence on $z$ and $\zeta$ (which is to be integrated over in (\ref{eq:FinalStateVectorFull}), (\ref{eq:FinalStateVectorExpansion})) has been circumvented by the device of formally introducing appropriate differentiations with respect to the $\xi' , \xi$ variables, so that (\ref{eq:FreeKernelExplicit})
normalised with reference to the noninteracting case, is sufficient provided that $K^{-1}$ is calculable.
For the case of $R$ constant this is easily checked to be
\begin{align}
\label{eq:InverseKernelSymbolic}
K^{-1}(t,t') = & \, - \theta(t-t') e^{(t-t')R} 
\end{align}
subject to $K^{-1}(t,t') = 0$ if $t \le t'$.

Consider then the noninteracting case (\ref{eq:FinalProbabilityComponent}), (\ref{eq:InverseKernelSymbolic}) with $S_1 \equiv 0$. Clearly the necessity to set the ancillary variables equal to zero \emph{after} differentiation means that the only viable initial probability state vector is one also with $L$ extant taxa, \emph{and} with identical momentum labels. Explicitly we have
\begin{align}
\label{eq:FreeProbabilityComponent}
& \, P_T^{\alpha_1 {\mathbf k}_1 \cdots \alpha_L {\mathbf k}_L} = 
        \frac{\partial}{\partial   i\xi'_{\alpha_1 {\mathbf k}_1}(T)} \cdots 
   \frac{\partial}{\partial i\xi'_{\alpha_L {\mathbf k}_L}(T)} \cdot   P_0^{\beta_1 {\mathbf k}_1 \cdots \beta_L {\mathbf k }_L}  
         \frac{\partial}{\partial \xi^{\beta_1 {\mathbf k}_1 }(0)} \cdots 
        \frac{\partial}{\partial \xi^{\beta_L {\mathbf k}_L }(0)} \cdot \nonumber \\        
         & \,    \left. \exp{  \int \hspace*{-.2cm} \int_0^T dt dt' \theta(t-t')
             \sum_{\gamma, \delta, {\mathbf m}} i \xi'_{\gamma  {\mathbf m} }(t) 
             {M_{(t-t'){\mathbf m}}^{\gamma } }_{\delta  }
             \xi^{\delta  {\mathbf m} }(t') }\right|_{z \equiv \xi \equiv \xi'=0}
\end{align}
Differentiations with respect to $i\xi'$, $\xi$ with the corresponding momentum labels must be paired, leading finally to 
\begin{align}
  P_T^{\alpha_1 {\mathbf k}_1 \cdots \alpha_L {\mathbf k}_L} = & \, 
\sum_{\delta_{i}} 
        { {M_{T{\mathbf k}_1}}^{\alpha_{1}} }_{\beta_{1}} 
     {{M_{T {\mathbf k}_2}}^{\alpha_{2}}}_{\beta_{2}} \cdots 
{{M_{T{\mathbf k}_L}}^{\alpha_{n}}}_{\beta_{n}}
P_0^{\beta_1 {\mathbf k}_1 \cdots \beta_L {\mathbf k }_L} 
 \end{align}
as was derived informally in (\ref{eq:PtensorFiniteEvolution}), (\ref{eq:PtensorFiniteEvolution2}).

Turning to the interacting case, we are interested in the final state probability for $L$ taxonomic units, assigned momenta ${\mathbf k}_1$, ${\mathbf k}_2$, $\cdots$$ {\mathbf k}_L$ say, arising from an initial state with one taxon (the root) with momentum
${\mathbf k}_0$, thus the probability component for a $L \leftarrow 1$ scattering process in the model. Once again, the necessity to set the ancillary variables equal to zero \emph{after} differentiation selects nonvanishing contributions corresponding to precisely degree $L-1$ in the power series expansion of the exponential of the interaction term 
${S_1[  \frac{\partial}{\partial {i\xi'}}, \frac{\partial}{\partial \xi  }} ] $ (see (\ref{eq:FinalProbabilityComponent}) and (\ref{eq:InteractionTermAction})): 
\begin{align}
\label{eq:IntProbabilityComponent}
 P_T^{\alpha_1 {\mathbf k}_1 \cdots \alpha_L {\mathbf k}_L} \!=\! 
& \, \frac{1}{(L\!-\!1)!}\left[\!-\! \int_{0}^{T} \hspace*{-8pt}{dt}\frac{(\!-\!i)^3}{2} \sum_{I}\sum_{{\mathbf k}, {\mathbf l},{\mathbf m}}
    \delta(t-t_{I})\delta({\mathbf k}\!-\!{\mathbf l}\!-\!{\mathbf m})
    {\frac{\partial}{\partial \xi^{{\mathbf l} \alpha}(t)} } {\frac{\partial}{\partial \xi^{{\mathbf m} \beta}(t)}}
    {\Gamma^{\alpha \beta}}_{\gamma}{\frac{\partial}{\partial \xi'_{{\mathbf m} \gamma}(t)}}  \right]^{(L\!-\!1)} \nonumber \\
      \cdot & \,  \frac{\partial}{\partial   i\xi'_{\alpha_1 {\mathbf k}_1}(T)} \cdots 
   \frac{\partial}{\partial i\xi'_{\alpha_L {\mathbf k}_L}(T)} \cdot   P_0^{\beta_0 {\mathbf k}_0}
         \frac{\partial}{\partial \xi^{\beta_0 {\mathbf k}_0 }(0)} \cdot \nonumber \\        
         & \,    \left. \exp{ - \left[ \int \hspace*{-6pt} \int_0^T dt dt' \theta(t-t')
             \sum_{\gamma, \delta, {\mathbf m}} i \xi'_{\gamma  {\mathbf m} }(t) 
             {M_{(t-t'){\mathbf m}}^{\gamma } }_{\delta  }
             \xi^{\delta  {\mathbf m} }(t')\right] } \right|_{z \equiv \xi \equiv \xi'=0}
\end{align}
It is convenient at this stage also to choose canonical momenta (binary $L$-vectors, with a scaling of $\pi$ understood) ${\mathbf k}_0=(1,1,\ldots,1)$ for the root, and ${\mathbf k}_1=(0,0,\ldots,1)$, ${\mathbf k}_2=(0,\ldots,1,0)$, $\cdots$ ${\mathbf k}_2=(0,\ldots,0,1)$ for the edges (or decimal equivalents $\vec{1}$, $\vec{2}$, $\vec{4}$, $\cdots$). For formal analysis with a specific tree, it may in fact be more
powerful to consider \textit{all} such admissible $L \leftarrow 1$
momentum routing schemes, however for combinatorial purposes any fixed assigment is sufficient.

For $L=2$ there is only one interaction, whose time is forced to be $t=t_1$.  Performing the differentiation of the exponential of the free kernel with respect to $\xi^{\beta_0 {\mathbf k}_0}(0)$ gives
\begin{align}
 P_T^{\alpha_1 {\mathbf k}_1 \alpha_2 {\mathbf k}_2} \!=\! 
 \left[ +\frac 12 \sum_{{\mathbf k}, {\mathbf l},{\mathbf m}}
    \delta({\mathbf k}\!-\!{\mathbf l}\!-\!{\mathbf m})
    {\frac{\partial}{\partial \xi^{{\mathbf l} \alpha}(t_1)} } 
 {\frac{\partial}{\partial \xi^{{\mathbf m} \beta}(t_1)}}
    {\Gamma^{\alpha \beta}}_{\gamma}  
 {\frac{\partial}{\partial  i \xi'_{{\mathbf k} \gamma}(t_1)}}  \right]  
  \cdot   \frac{\partial}{\partial   i\xi'_{\alpha_1 {\mathbf k}_1}(T)}  
   \frac{\partial}{\partial i\xi'_{\alpha_2 {\mathbf k}_2}(T)} \cdot  & 
   \nonumber \\
 \left[   \!+\! \int_{0}^{T} \hspace*{-6pt}{dt} \, i\xi'_{\lambda {\mathbf k}_0}(t) 
M_{t {\mathbf k}_0} {\mbox{}^\lambda}_{\beta_0}   \right] P_0^{\beta_0 {\mathbf k}_0}\cdot 
          \left. \exp{ + \left[ \int \hspace*{-7pt} \int_0^T dt dt' \theta(t-t') 
             \sum_{\gamma, \delta, {\mathbf m}} i \xi'_{\rho  {\mathbf m} }(t)        
             (M_{(t-t'){\mathbf m}} {\mbox{})^{\rho }}_{\sigma} 
             \xi^{\sigma  {\mathbf m} }(t')\right] } \right|_{ \xi \equiv 0 \equiv \xi'} &  \nonumber
\end{align}
and carrying out the two remaining $\xi$ differentiations leads to
\begin{align}
 P_T^{\alpha_1 {\mathbf k}_1 \alpha_2 {\mathbf k}_2} \!=\! 
 \left[ +\frac 12 \sum_{{\mathbf k}, {\mathbf l},{\mathbf m}}
    \delta({\mathbf k}\!-\!{\mathbf l}\!-\!{\mathbf m})
    {\Gamma^{\alpha \beta}}_{\gamma}  
{ \frac{\partial}{\partial   i\xi'_{\alpha_1 {\mathbf k}_1} (T)}}  
  { \frac{\partial}{\partial i\xi'_{\alpha_2 {\mathbf k}_2}(T)}}{\frac{\partial}{\partial i \xi'_{{\mathbf k} \gamma}(t_1)}}   \right] \cdot & 
   \nonumber \\
 \left[   \!+\! \int_{t_1}^{T} \hspace*{-6pt}{dt} \, i\xi'_{\lambda {\mathbf l} }(t) 
M_{t {\mathbf l}} {\mbox{}^\lambda}_{\alpha}   \right] 
 \left[   \!+\! \int_{t_1}^{T} \hspace*{-6pt}{dt} \, i\xi'_{\mu {\mathbf m}}(t) 
M_{t {\mathbf m}} {\mbox{}^\mu}_{\beta}   \right] 
 \left[   \!+\! \int_{0}^{T} \hspace*{-6pt}{dt} \, i\xi'_{\nu{\mathbf k}_0}(t) 
M_{t {\mathbf k}_0} {\mbox{}^\nu}_{\beta_0}   \right] P_0^{\beta_0 {\mathbf k}_0}\cdot &  \nonumber \\
          \left. \exp{+ \left[ \int \hspace*{-7pt} \int_0^T dt dt' \theta(t-t') 
             \sum_{\gamma, \delta, {\mathbf m}} i \xi'_{\rho  {\mathbf m} }(t)        
             M_{(t-t'){\mathbf m}} {\mbox{}^{\rho }}_{\sigma} 
             \xi^{\sigma  {\mathbf m} }(t')\right] } \right|_{ \xi \equiv 0 \equiv \xi'} &  \nonumber
\end{align}
For a nonzero result the remaining $\xi'(t_1)$ and two $\xi'(T)$ differentiations can only be applied to the terms standing in front of the exponential. Moreover, the implicit $\theta$ terms require the $\xi'(t_1)$ differentiation to be applied only to the ${\mathbf k}_0$ integral, thus fixing ${\mathbf k} = {\mathbf k}_0$. Finally since ${\mathbf k}_0 = {\mathbf k}_1 + {\mathbf k}_2 = {\mathbf l} + {\mathbf m}$ there are two equivalent ways to apply the remaining differentiations (cancelling the symmetry factor $\frac 12$ in the interaction term) giving finally
\begin{align}
\label{eq:IntProbabilityComponentLeq2}
 P_T^{\alpha_1 {\mathbf k}_1 \alpha_2 {\mathbf k}_2} \!=\! & \,
{\Gamma^{\alpha \beta}}_{\gamma}  
(M_{(T-t_1) {\mathbf k}_1}){\mbox{}^{\alpha_{1}}}_{\alpha} 
(M_{(T-t_1) {\mathbf k}_2}) {\mbox{}^{\alpha_2}}_{\beta}  
 (M_{t_1 {\mathbf k}_0}){\mbox{}^\gamma}_{\beta_0} P_0^{\beta_0 {\mathbf k}_0}. 
\end{align}
In the general case systematic diagrammatical rules (Feynman rules) can easily be ascribed and tabulated for the evaluation of 
(\ref{eq:IntProbabilityComponent}). On the basis of the above $L=2$ (first order) case and similar considerations for $L=3$ (second order), 
all possible probability component contributions for $L$ taxa are constructed from the matrix element for  $L \leftarrow 1$ scattering as follows:
\subsection*{Feynman rules for phylogenetic trees}
\begin{itemize}
\item[1] Diagrams consist of $2L\!-\!1$ directed edges, $L\!-\!1$ vertices with internal nodes, one external root and $L$ leaf nodes;
\item[2] To each element is assigned character and momentum labels as in table \ref{tab:FeynRules}; specifically,
\item[3] Root and leaf edge momenta are assigned canonical binary $L$-vectors (see text); momentum conservation between ingoing and outgoing edge momenta is imposed.
\item[4] Vertices (internal nodes) are assigned interaction times ordered $t_1 < t_2 <  \cdots t_{L\!-\!1}$;
\item[5] The root node is assigned time $t=0 =t_0$, and the leaves are assigned time $t=T=t_L$.
\end{itemize}
To these labelled diagrammatical elements, the following algebraic terms are associated:
\begin{itemize}
\item[6] For each directed edge, a Markov transition matrix for time interval $\Delta = (t_{I'}-t_I)$, $0 \le I \le L\!-\!1$ between the target and source nodes, and for its assigned edge momentum, and matrix element determined by the source and target character labels assigned (see table \ref{tab:FeynRules});
\item[7] To each vertex, an appropriate component of the $\Gamma$ structure coefficient (see table \ref{tab:FeynRules});
\item[8] Consistent combinations of these elements, summed over all internal momenta and character indices, with valid momentum conservation, correspond to contributions from all possible labelled $L$ leaf binary trees.
\end{itemize}
\begin{table}
\begin{tabular}{rcl}
\hline
  Element &  & Term \\
  \hline
  {\raisebox{.6cm}{internal edge}} &  \hspace*{10pt}
\begin{pspicture}(2,2)
\psset{framearc=0.0,framesep=0.0truecm,nodesep=8pt,unit=0.75truecm,
      arrowsize=4pt 2,arrowinset=0.4}
\def\pstlw{0.8pt}      
\psline{-}(1,1.8)(1,.2)
\psline{->}(1,1.8)(1,.8)
\psline[linestyle=dashed, linewidth=.5pt]{-}(.5,1.8)(1,1.8)
\psline[linestyle=dashed, linewidth=.5pt]{-}(.5,.2)(1,.2)
\rput(0,1.8){$t_I$}
\rput(0,.2){$t_{I'}$}
\rput(.7,1.1){$\mathbf k$}
\rput(1.4, 1.8){$\beta_{\mathbf k}$}
\rput(1.4, .2){$\alpha_{\mathbf k}$}
\end{pspicture}
 &  {\raisebox{.6cm}{${(M_{\Delta_{\mathbf k} {\mathbf k})}\mbox{}^{\alpha_{\mathbf k}}}_{\beta_{\mathbf k}}$}}
 \\
{\raisebox{.6cm}{root}} & \hspace*{10pt}
\begin{pspicture}(2,2)
\psset{framearc=0.0,framesep=0.0truecm,nodesep=8pt,unit=0.75truecm,
      arrowsize=4pt 2,arrowinset=0.4}
\def\pstlw{0.8pt}      
\psline{-}(1,1.8)(1,.2)
\psline{->}(1,1.8)(1,.8)
\psline[linestyle=dashed, linewidth=.5pt]{-}(.5,1.8)(1,1.8)
\psline[linestyle=dashed, linewidth=.5pt]{-}(.5,.2)(1,.2)
\rput(.1,1.8){$0$}
\rput(.1,.2){$t_1$}
\rput(.6,1.1){${\mathbf k}_0$}
\rput(1.55 ,1.8){$\beta_{{\mathbf k}_0}$}
\rput(1.55, .2){$\alpha_{{\mathbf k}_0}$}
\pscircle[linewidth=0.8pt,fillstyle=solid,fillcolor=white](1,1.8){.14} 
\end{pspicture}
 &  \raisebox{.6cm}{${(M_{\Delta_0 {\mathbf k}_0})
 \mbox{}^{\alpha_{{\mathbf k}_0}}  }_{\beta_{{\mathbf k}_0}}{p_0}^{\beta_{{\mathbf k}_0}} $}  
 \\
{\raisebox{.6cm}{leaf}}  & \hspace*{10pt} 
\begin{pspicture}(2,2)
\psset{framearc=0.0,framesep=0.0truecm,nodesep=8pt,unit=0.75truecm,
      arrowsize=4pt 2,arrowinset=0.4}
\def\pstlw{0.8pt}      
\psline{-}(1,1.8)(1,.2)
\psline{->}(1,1.8)(1,.8)
\psline[linestyle=dashed, linewidth=.5pt]{-}(.5,1.8)(1,1.8)
\psline[linestyle=dashed, linewidth=.5pt]{-}(.5,.2)(1,.2)
\rput(.1,1.8){$t_{J}$}
\rput(.1,.2){$T$}
\rput(.6,1.1){${\mathbf k}_i$}
\rput(1.55 ,1.8){$\beta_{{\mathbf k}_i}$}
\rput(1.55, .2){$\alpha_{{\mathbf k}_i}$}
\pscircle[linewidth=0.8pt,fillstyle=solid,fillcolor=white](1,.2){.14} 
\end{pspicture}
&  \raisebox{.6cm}{${(M_{\Delta_{i} {\mathbf k}_i})
 \mbox{}^{\alpha_{{\mathbf k}_i}}  }_{\beta_{{\mathbf k}_i}}$}  
  
 \\
{\raisebox{.6cm}{vertex}}  & \hspace*{10pt} 
\begin{pspicture}(2,2)
\psset{framearc=0.0,framesep=0.0truecm,nodesep=8pt,unit=0.75truecm,
      arrowsize=4pt 2,arrowinset=0.4}
\def\pstlw{0.8pt}      
\psline{-}(1,1.8)(1,1)
\psline{->}(1,1.8)(1,1.2)
\psline{-}(1,1)(.3,.2)
\psline{-}(1,1)(1.7,.2)
\psline{->}(1,1)(.58,.52)
\psline{->}(1,1)(1.42,.52)
\rput(.7,1.8){$\mathbf k$}
\rput(1.45,1.8){$\alpha_{\mathbf k}$}
\rput(.4,.7){$\mathbf l$}
\rput(0,.2){$\beta_{\mathbf l}$}
\rput(1.7,.7){$\mathbf m$}
\rput(2.2,.2){$\beta_{\mathbf m}$}
\pscircle[linewidth=0.8pt,fillstyle=solid,fillcolor=white](1,1){.14} 
\end{pspicture}
 & \raisebox{.6cm}{$ {\Gamma^{\alpha_{\mathbf k}}}_{\beta_{\mathbf  l} \beta_{\mathbf m}} 
 \delta({\mathbf  k} - {\mathbf l} -{\mathbf m})$}
  \\
\hline
\end{tabular}
 \centering 
 \caption{Feynman rules for evaluating probabilities for $L \leftarrow 1$ scattering in phylogenetic branching model.
 Trees are a combination of labelled graphical elements. Each tree contributes a term to the total likelihood or pattern frequency.
 ${(M_{\Delta {\mathbf k}})\mbox{}^{\alpha}}_{\beta}$ is t the Markov transition matrix for edge $\mathbf k$ and time interval (edge length) $\Delta$, and ${\Gamma^\alpha}_{\beta \beta'}$ is the vertex structure coefficient ($\equiv {\delta^\alpha}_\beta {\delta^\alpha}_{\beta'}$). See concluding remarks for comments on the role of the group of time reparametrisations.}
 \label{tab:FeynRules}
\end{table}

Using these rules, likelihoods can thus be written down autonomously and diagrammatically, without reference to the path integral context; however as stressed in the main text, the utility of the path integral formulation is precisely to provide a self-contained prescription for them without the explicit need to enumerate trees. 

\end{appendix}

\newpage

\noindent
{\sc 
James D. Bashford$^{1}$,\\
Peter D. Jarvis$^{2}$, \\  
Jeremy G. Sumner$^{3}$,\\
University of Tasmania, School of Mathematics and Physics,
GPO Box 252C, \\ 7001 Hobart, TAS, Australia. \\[.3 cm]
$^{1}$ Australian Postdoctoral Fellow, {\small\tt 
James.Bashford@utas.edu.au} \\
$^{2}$ Alexander von Humboldt Fellow, {\small\tt 
Peter.Jarvis@utas.edu.au} \\
$^{3}$ Australian Postgraduate Award, {\small\tt 
Jeremy.Sumner@utas.edu.au}
\\[1cm]
Report Number UTAS-PHYS-2004- }
\vfill
\tableofcontents
\end{document}